%% file: Kagome_topo.tex
\documentclass[12pt]{iopart}

\usepackage{graphicx}
\usepackage{color}
\usepackage{ulem}
\begin{document}

\title{Classical Topological Order in Kagome Ice}

\author{Andrew J. Macdonald,$^1$
Peter C. W. Holdsworth,$^2$
Roger G. Melko$^1$}

\address{$^1$Department of Physics and Astronomy, University of Waterloo, Ontario, N2L 3G1, Canada}
\address{$^2$Laboratoire de Physique, Ecole Normale Superieure de Lyon, Universite de Lyon, CNRS, 46 Allee dÕItalie, 69364 Lyon Cedex 07, France}

\begin{abstract}
We examine the onset of classical topological order in a nearest-neighbor kagome ice model.  Using Monte Carlo simulations, we characterize the topological sectors of the groundstate using a non-local cut measure which circumscribes the toroidal geometry of the simulation cell.  We demonstrate that simulations which employ global loop updates that are allowed to wind around the periodic boundaries cause the topological sector to fluctuate, while restricted local loop updates freeze the simulation into one topological sector.  The freezing into one topological sector can also be observed in the susceptibility of the real magnetic spin vectors projected onto the kagome plane.  The ability of the susceptibility to distinguish between fluctuating and non-fluctuating topological sectors should motivate its use as a local probe of topological order in a variety of related model and experimental systems.
\end{abstract}

\section{Introduction}

There has been great interest recently in quantum liquids which fail to find symmetry-broken states down to temperatures far below that set by the leading energy scale.  Rather, such systems occupy an extensive band of low lying states which retain some of the fluid-like features of high-temperature phases even in the absence of configurational disorder. Magnetic systems have become the canonical systems for study and this disordered state is referred to as a quantum spin liquid \cite{Anderson,LeonSL}.  Spin liquids, however, differ in important ways from their higher-temperature paramagnetic counterparts, due to the existence of subtle correlations resulting from highly-constrained collective motion of the magnetic moments over the low energy manifold of states. The constraints can lead to hidden topological information encoded into the low energy states, and 
quantum spin liquids can be classified by this topological order buried in their disordered  quantum groundstates \cite{Wen}.

The concept of topological order  can also be applied to classical systems \cite{Henley,Claudio}.  A classical system with topological order is characterized by a highly degenerate manifold of ground states whose configurations are subject to a local topological constraint. Imposing the constraint everywhere leads again to information being encoded non-locally into each configuration, even in the absence of symmetry breaking, or selection of any particular state down to zero temperature.  
In this paper, we examine a concrete and experimentally realizable example of a topologically constrained system, that of  ``kagome ice'' \cite{Kice2,Kice1}.  This phase can be realized to a good approximation in spin ice materials such as Dy$_2$Ti$_2$O$_7$ by placing a magnetic field of intermediate strength along the [111] crystallographic direction.  With this field configuration a magnetization plateau occurs that corresponds to a dimensional reduction of the magnetic degrees of freedom into the kagome planes perpendicular to the field  \cite{Isakov}.  Application of the ice rules for each tetrahedral unit of spins is equivalent  to applying a local topological constraint to the perpendicular, in-plane degrees of freedom. Within the approximation of nearest neighbour interactions only, the resulting degenerate manifold of states is identical to two systems: hard core dimers close-packed on a honeycomb lattice; and, the groundstate of an antiferromagnetic Ising spin model on the kagome lattice in the presence of a magnetic field. 
Such discrete systems, subject to a local constraint, are characterized by an extensive ground state degeneracy and a gap for excitations out of the constrained manifold.
These excitations break the constraints leading to the proliferation of deconfined topological defects \cite{Castelnovo08}.  In this paper, we define and review several measures of topological order through a topological {\it number} or {\it sector}.  We then use Monte Carlo simulations to study the fluctuations of these quantities and the onset of topological order in the kagome ice model. Using local and non-local updates, we examine the onset of ergodicity-breaking between different topological sectors as one cools into the groundstate phase.
We demonstrate that one can discern the ergodicity breaking corresponding to the onset of topological order using the magnetic susceptibility  -- a local quantity that is easily measurable and accessible in principle in experiments. We show that topological order occurs here uniquely through the application of constrained dynamics, which limits collective ``loop'' moves to clusters of microscopic length. We compare our results with a collective paramagnetic system \cite{Villain} without topological constraints, so called ``Wills-ice'' \cite{Wills}, whose ground state manifold maps onto the complete low energy phase space of the Ising antiferromagnet on the kagome lattice.

\section{Pseudospin Hamiltonian and Topological Sectors}

\begin{figure}[b!] 
\begin{center}
\includegraphics[width=3.1in]{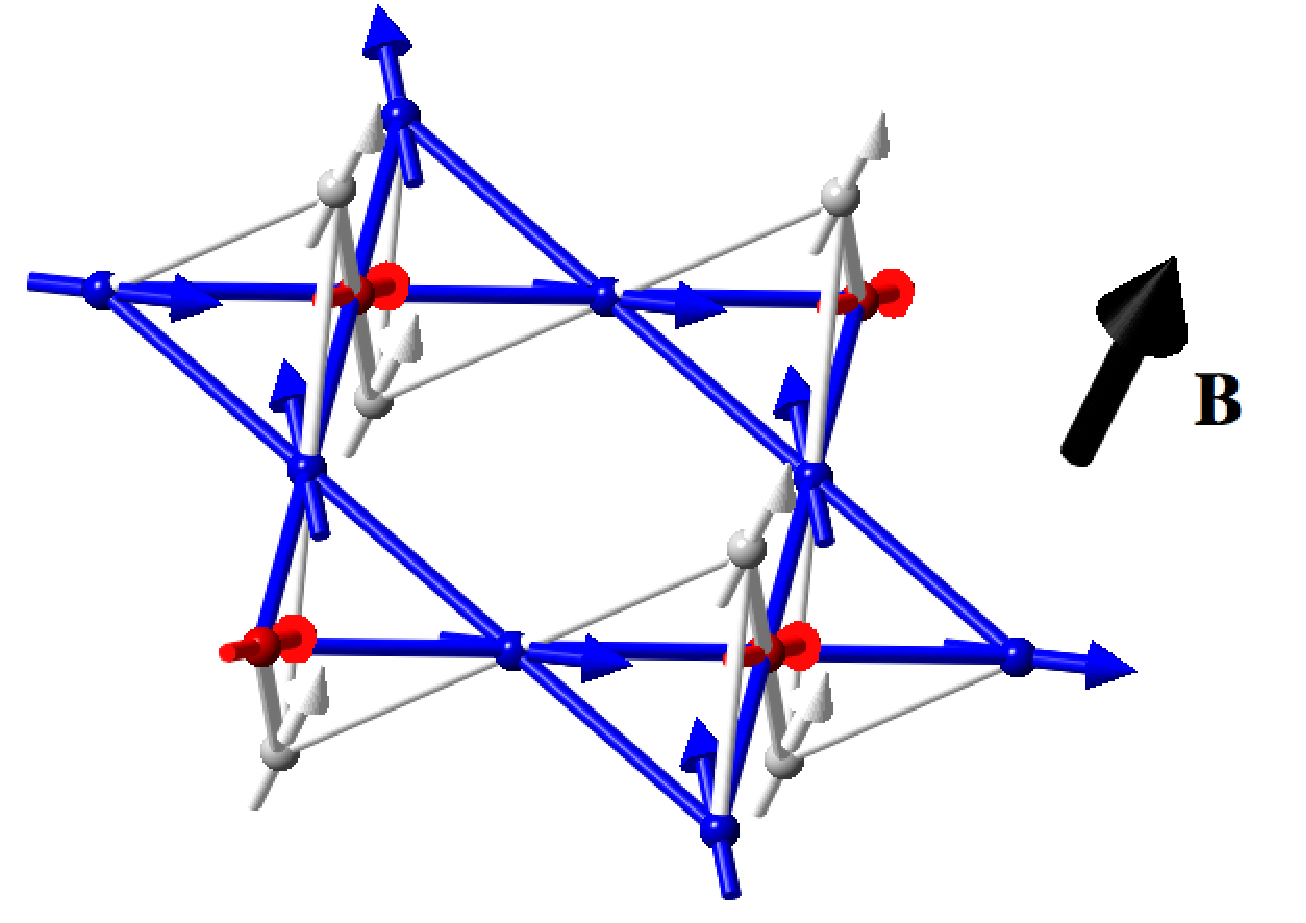}\label{F1a}
\includegraphics[width=2.8in]{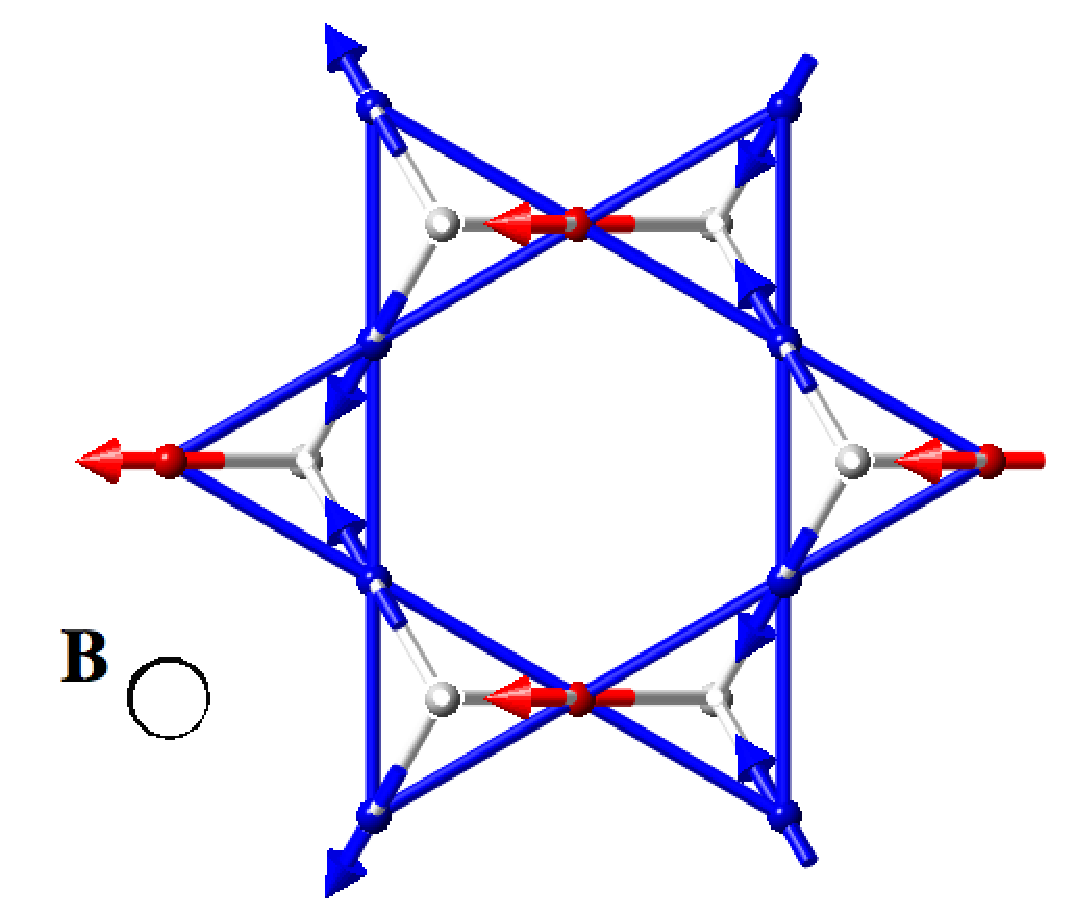}\label{F1b}
\caption{A section of the kagome lattice (blue) corresponds to a 2D plane in the lattice of corner sharing tetrahedra.  The white spins are assumed to be static, pinned by a [111] external field, and therefore do not represent magnetic degrees of freedom in the model Equation~(\ref{Ham}).  Rather, they are intended to motivate the internal field term $h$ affecting the kagome spin degrees of freedom.  In the spin configuration shown, minority spins are shown in red.}
%Two snapshots of the system, with the kagome lattice plane in blue, the minority spins in red, and the out of plane spins pinned by the field shown in grey.}
\label{3Dfig}
\end{center}
\end{figure}

We begin with the Ising antiferromagnet on the kagome lattice with Hamiltonian:
\begin{equation}
H = J \sum_{\langle ij \rangle} \sigma_i \sigma_j - h \sum_{i} \sigma_i \label{Ham}
\end{equation}
where we refer  to  the $\sigma_i = \pm 1$ as pseudospin variables and where $h$ is an external field in reduced units.  The kagome lattice is built of a triangular Bravais lattice with a three spin basis which we take to be the spins on a ``left'' triangle (see Figure~1). In zero field the best this highly frustrated system can do in lowering its energy is to assure two spins up and one down on each triangle ($uud$)- or the inverse ($ddu$). This leads to an extensive ground state entropy, $S/N=0.502$ ${\rm k_B}$ \cite{Kagome502},   and although the configurations on each triangle are not independent, all six combinations of $uud$ and $ddu$ figure in the ground state manifold. Applying a magnetic field breaks the $Z_2$ symmetry of $uud$ and $ddu$ states but still does not lead to an ordered groundstate. Rather, the groundstate entropy is reduced, but remains finite at $S/N=0.108$ ${\rm k_B}$ \cite{IsingFrust}. One can make a direct mapping between the antiferromagnet and a 2D ice-like system by defining in-plane spin degrees of freedom lying along the axes joining the triangles of the kagome lattice. 
The mapping between the real spins  ${\bf S}_i$ and pseduospins $\sigma_i$ is
\begin{eqnarray}
{\bf S}_i = \hat{d}_i \sigma_i
\end{eqnarray}
where $\hat{d}_i$ are the three unit vectors joining a left triangle to a right triangle in Figure~1,
\begin{eqnarray}
\hat{d}_0 = \left[{-\frac{\sqrt 3}{2} , -\frac{1}{2} }\right], \ \
\hat{d}_1 = \left[{\frac{\sqrt 3}{2}, -\frac{1}{2}}\right], \ \
\hat{d}_2 = \left[{0, 1}\right] ,
\label{eq:map}
\end{eqnarray}
which leads to a frustrated ferromagnetic Hamiltonian whose exchange term reads
\begin{equation}
H = - 2 J \sum_{\langle ij \rangle} {\bf S}_i \cdot {\bf S}_j  \label{Ham2}.
\end{equation}
Within the real spin description, the reduced manifold of states, with broken $Z_2$ symmetry corresponds to either two spins pointing out of and one into each left triangle (see the Appendix, Figure~9), or vise versa, while the phase space with full symmetry corresponds to a mixture of both. Consider the case with reduced symmetry, with two-out and one-in and define an effective field ${\bf B}_i$ such that
${\bf B}_i = 2 {\bf S}_i$ for a spin entering a left triangle (and leaving a right triangle) and that ${\bf B}_i={\bf S}_i$ for all other spins. The constraint imposes that the total flux of ${\bf B}$ into each triangle is zero; that is, that the ground state condition plus broken $Z_2$ symmetry corresponds to a discrete, divergence free condition,  $\vec \nabla \cdot {\bf B}=0$. Breaking this symmetry thus leads to an emergent U(1) gauge field encoded into each disordered ground state configuration \cite{U1}. The constrained system also maps onto a close-packed dimer system on a honeycomb lattice by placing a dimer along the line joining the centres of each pair of triangles connected by a minority spin \cite{Moessner03}. Equivalent gauge field descriptions have been made in other systems with dimers tiling a bipartite lattice \cite{Huse03}. Through the emergent gauge field, excitations that break the constraints decompose into pairs of quasi-particles carrying topological charge.  It has recently been shown, in the  related physics of three dimensional (3D) spin ice materials,  that in addition to the topological properties the charges interact via an effective Coulomb law, making them examples of magnetic monopole quasi-particles \cite{Castelnovo08}. 

The name ``kagome ice'' was first given to a magnetic ice model interacting via Hamiltonian (\ref{Ham}), with access to the complete phase space of low energy states ($h=0$) by Wills {\it et.~al.}~\cite{Wills}. We refer to this system as ``Wills-ice'' to distinguish it from the constrained manifold of states with broken $Z_2$ symmetry, attained by applying a magnetic field along the $[111]$ axis of a 3D model system, the nearest neighbour spin ice \cite{Isakov,Moessner03}. In this model, the spins sit on the vertices of a pyrochlore lattice of corner sharing tetrahedra and lie along axes joining the centres of the tetrahedra, in analogy with the two dimensional (2D) spins described above. The field isolates kagome planes perpendicular to the $[111]$ plane pinning the central spins of the tetrahedra, as shown in Figure 1. The 3D spins are subject to the ice rules with two spins in and two out of each unit, leaving the other three spins on each tetrahedra lying on a triangular element of a kagome lattice, with projections in the perpendicular plane.  The 3D ice rules mean that the projected spin components are constrained to be two-in, one-out on each triangle, or the inverse, with the choice dictated by the magnetic field direction along $\pm [111]$. While the in-plane components map exactly onto our ``real spins'' in Hamiltonian (\ref{Ham2}), the components parallel to $[111]$ are equivalent to the pseudo spins \cite{Moessner03}. We simulate this constraint by applying a field $h=2J$ to the pseudo spin Hamiltonian, which corresponds to the effective field of a static fourth spin with tetrahedral geometry, associated with each triangle. $h$ fixes the energy scale for topological defects and the energy scale on which these topological ice rules are obeyed. In what follows we shall refer to the constrained kagome ice as ``$[111]$ ice''.

\begin{figure}[b!] 
\begin{center}
\includegraphics[width=5in]{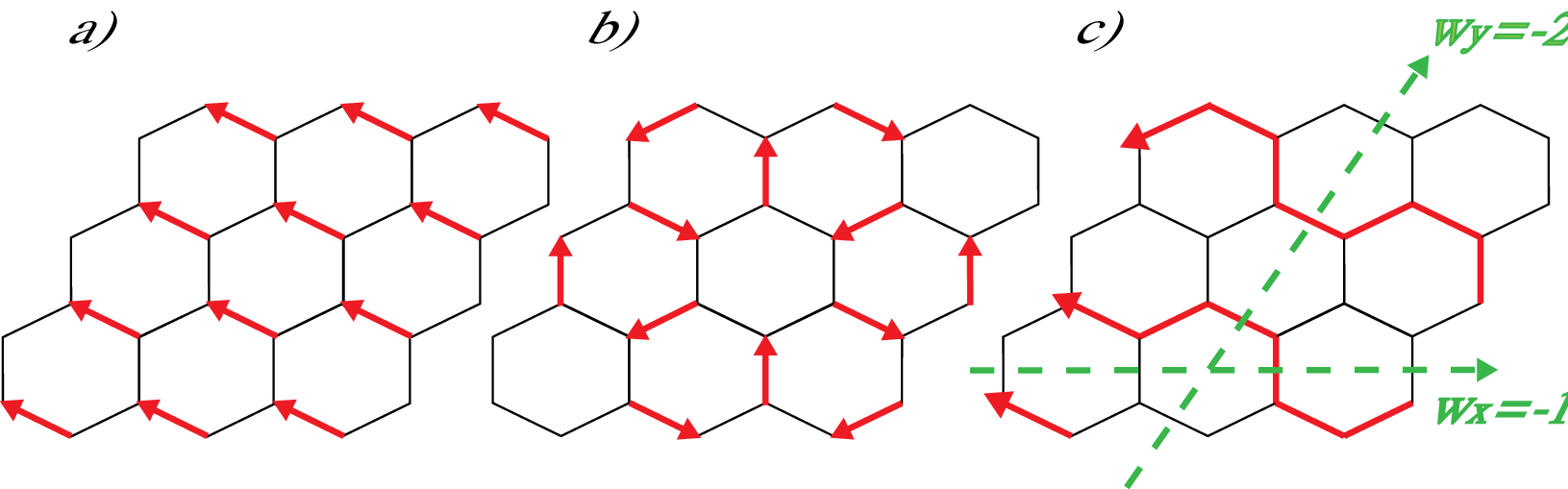}
\caption{Two dimer coverings of the dual honeycomb lattice possible under the ice rules, (a) shows the state of interest $| C \rangle$, (b) shows the reference columnar state $| C^{\prime} \rangle$ and (c) shows their associated transition graph with winding numbers.}
\label{Winding}
\end{center}
\end{figure}

The consequences of the topological constraints have previously been discussed in the context of
hard-core dimer coverings on the honeycomb lattice.  This order can be characterized in several ways, the most well-known being through the use of winding numbers, introduced by Rokshar and Kivelson on the frustrated square lattice \cite{Rokshar}.  Their method extends to all bipartite lattices, and is defined via construction of a transition graph between any hard-core dimer covering, $| C \rangle$, and a chosen reference state $| C^{\prime} \rangle$. 
Given that the honeycomb lattice is bipartite, it is possible to orient the dimers of $| C \rangle$ from the A to B sub-lattice. Orienting the dimers of $| C^{\prime} \rangle$ in the opposite direction and then superimposing the two states produces a transition graph consisting of  non-interacting loops (see Figure \ref{Winding}).  On a two-torus the topological sector of $| C \rangle$ can then be characterized by the winding numbers $w_x$,$w_y$, defined as the net number of clockwise loops minus the net number of counterclockwise loops encircling the torus in the x and y direction respectively. These winding numbers are bound between $-L \leq w_x, w_y \leq L$, where L is the linear system size of the underlying kagome lattice.  Note that loops of microscopic length (smaller than the system size) contribute zero to the winding number, as a line circumscribing the boundary cuts all such loops twice, circulating once in each direction. 
 
An alternative way to determine the winding numbers of a particular configuration is to avoid constructing the transition graph or defining a reference state -- taking instead the measurement directly on the spins of the kagome lattice.  This can be done by considering closed ``cuts'' along x and y respectively, and counting the number of minority elements - dimers within the dimer picture; minority pseudo spins, or real spins in the magnetic picture.  As we apply a field to the pseudo spins in the ``up'' direction, the minority spins are ``down'' (real spins are ``out'' of the ``up'' triangles in Figure 3). The cut number, $Hx_{| C \rangle}, Hy_{| C \rangle}$, is then defined as the number of down pseudo spins, or out real spins encountered along the $x$ and $y$ directions, as shown in Figure 3.
As these spins indicate the position of dimers on the honeycomb lattice, cut numbers can be constructed by adding $+1$ every time a dimer is cut along the path. This definition gives only positive numbers, as it corresponds to the $Z_2$ symmetry being broken in a particular way. Placing the field in the opposite direction would isolate the opposite half of phase space and would require the definition of negative numbers.  Counting the number of down pseudo-spins means that the cut numbers remain valid also for unconstrained Wills ice; however the down pseudo-spins are not necessarily minority and the probability of encountering one along any path approaches $1/2$.

\begin{figure}%[h!] 
\begin{center}
\includegraphics[width=3.5in]{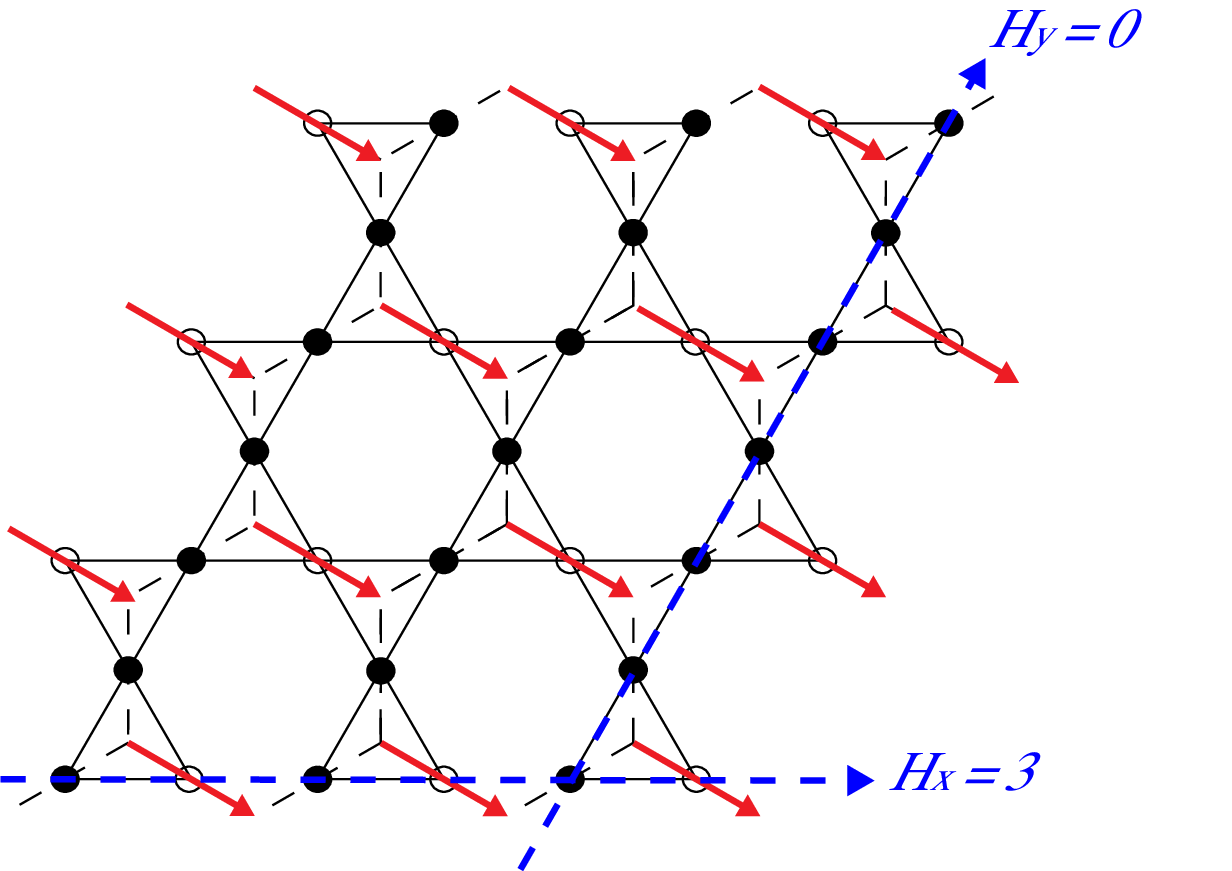}
\caption{Two possible cuts taken along the original kagome lattice with the honeycomb dimer lattice superimposed on top. Unfilled circles represent minority spins on the original kagome lattice, while the red arrows indicate dimers oriented from the A to B sub-lattices within the honeycomb lattice.}
\label{Kagome}
\end{center}
\end{figure}

To make this measure equivalent to the winding numbers the constrained [111] ice, as previously defined, one must take the cut numbers of the current state and subtract the cut numbers of the reference state (Equations \ref{eq:convertx}~-~\ref{eq:converty}). Note that there is a key difference between this treatment on the honeycomb lattice and that on the square lattice: on the honeycomb there is no possible reference state with cut numbers identically equal to zero, while on the square lattice using  columnar state as the reference, $Hx_{| C^{\prime} \rangle}, Hy_{| C^{\prime} \rangle} = 0$, so $Hx_{| C \rangle}, Hy_{| C \rangle} = w_x, w_y$ for every possible dimer covering.

\begin{equation}
w_x =  Hx_{| C \rangle} - Hx_{| C^{\prime} \rangle}  \\
\label{eq:convertx}
\end{equation}
\begin{equation}
w_y = Hy_{| C \rangle} - Hy_{| C^{\prime} \rangle}
\label{eq:converty}
\end{equation}

\begin{figure}%[h!] 
\begin{center}
\includegraphics[width=5in]{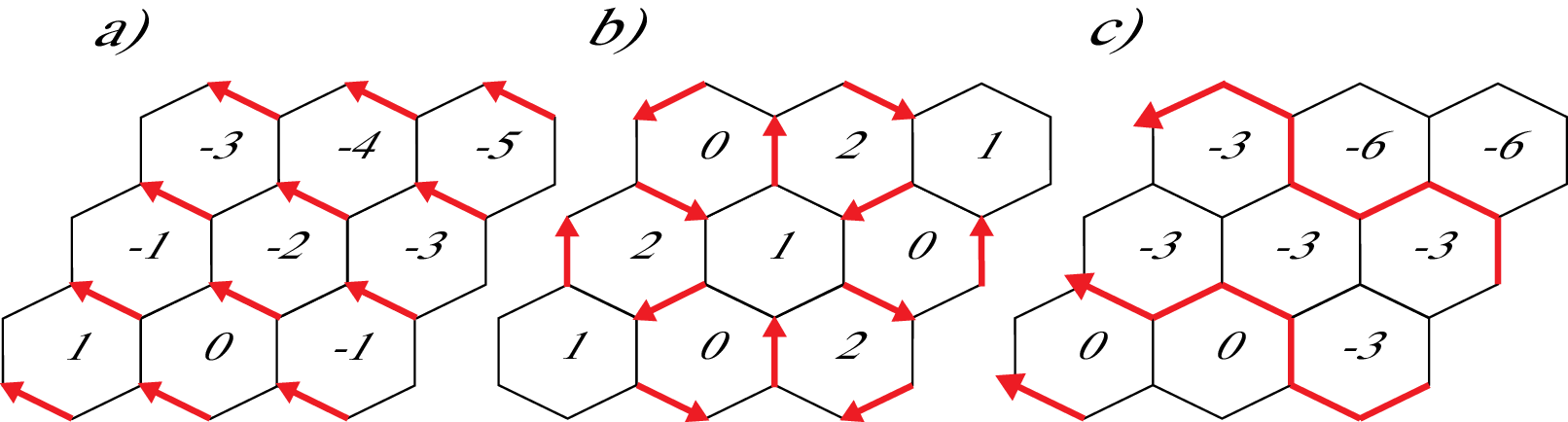}
\caption{Height model covering of the a) state $| C \rangle$, b) the reference state  $| C^{\prime} \rangle$ and c) of their transition graph. The difference in heights is constant within each loop.}
\label{Height}
\end{center}
\end{figure}

A third possibility for the definition of the topological sector is the construction of  a height model  on a bipartite lattice. Given the rules developed by Henley \cite{Henley}, a function $h(x,y)$ can be constructed to assign an integer height to each vertex on the lattice dual to the one on which the dimers reside.  The construction occurs as follows: set the height to be zero at an arbitrary plaquette, then assign integer heights by turning clockwise around the sites of the A sublattice, adding $+1$ when crossing a bond with no dimer and subtracting $z-1$ when crossing a dimer. Here $z$ is the coordination number of the dimer occupied lattice.
For the honeycomb lattice the dual lattice is the triangular lattice and assigning heights to its vertices amounts to assigning integer values to each plaquette of the honeycomb as shown in Figure \ref{Height}.   

If the height mapping is constructed for both the dimerization of interest and for the reference state as in Figure \ref{Height} then one can examine the difference in heights $\delta h = h_{| C \rangle} - h_{| C^{\prime} \rangle} $ mapped on the transition graph, for which the following holds: $\delta h$ is constant inside each loop of the transition graph,   $\delta h$ changes by $+z$ when crossing a clockwise loop of the transition graph, and the winding numbers correspond to the average height difference between both sides of the sample.  This last statement allows recovery of the winding numbers via the height model by taking the difference of heights at the edges of the lattice and dividing by the coordination number (Equations \ref{eq:heightx} - \ref{eq:heighty}).  For a lattice with periodic boundary conditions the height difference is taken between what the next height on the lattice should be and what it is upon wrapping around the boundary:

\begin{equation}
\delta h_x = h(2L,y) - h(0,y), \ \ \ w_x = \frac{\delta h_x}{z},  \\
\label{eq:heightx}
\end{equation}
\begin{equation}
\delta h_y = h(x,2L) - h(x,0), \ \ \ w_y = \frac{\delta h_y}{z}.
\label{eq:heighty}
\end{equation}

Thus, all three methods of characterizing the topology of the groundstate (transition graphs, cut numbers, and height mappings) are related.  In a numerical simulation, it is therefore only necessary to measure one in order to define the topological sector.  In the rest of this paper, we employ cut numbers measured directly on the kagome lattice, since this avoids the need for constructing the dimer model on the honeycomb lattice. %or its dual triangular lattice.  
In the next section, we compare measurements of the topological sectors defined via the cut number to measurements of susceptibility using a Monte Carlo simulation procedure.

\section{Classical Monte Carlo and Measurements of Susceptibility}

We perform Monte Carlo simulations on the pseudo-spin kagome Ising Hamiltonian Equation (\ref{Ham}).  We used both a loop update algorithm \cite{Loops,JPCreview} (adapted to the problem at hand) and a standard Metropolis algorithm allowing single spin flips.
  The loop update begins by first choosing an ``up'' ($\sigma_i = 1$) pseudospin at random; the loop ``head'' then enters a triangle associated with this pseudospin, and exits through a ``down'' ($\sigma_i=-1$) pseudospin.  The algorithm is adapted to work both in the constrained (Wills) and unconstrained [111] cases. In both situations the algorithm searches for a pathway flipping up and down pseudospins sequentially. When a minority spin is to be flipped, the algorithm has no choice and progresses deterministically to the next triangle. For a majority spin, it chooses one of the two directions at random. For the unconstrained case, majority and minority spins change at random along the loop, while for the constrained case they are fixed everywhere by the broken $Z_2$ symmetry.  
Since, for both cases the ground state consists of an extensive manifold of degenerate states, the loop algorithm works at $T=0$ where there is no internal energy, and where the specific heat goes to zero allowing us to assure ergodicity, or partial ergodicity among the low energy states.  At $T>0$, excitations appear above the degenerate manifold, leading to triangles which violate the local ice rules. If the loop encounters such a triangle it aborts. Metropolis updates in conjunction with the loop moves ensures a stochastic evolution of the system in this case.
 
We further allow two variations of the loop building algorithm. In the first we allow the loops to grow to any size including loops that span the periodic boundaries. We refer to this  as the ``long'' loop algorithm.  As demonstrated below, the loops that wind around the periodic boundaries of the system allow fluctuations of the topological sector.  We also define a ``short'' loop, which is built using the same algorithm, however if the loop size exceeds some pre-determined number of sites $N_{\rm loop}^{\rm max}$, the loop is aborted and the update is not performed.  Detailed balance is not affected if this procedure is followed.  Typically, we chose $N_{\rm loop}^{\rm max} < L$ where $L$ is the linear size of the lattice (of $N=L \times L \times 3$ spins), in order to study the effects of short loops on the ergodicity between topological sectors.

In addition to the usual thermodynamic estimators (such as energy and specific heat), we directly measure the topological cut number defined in the previous section (Figure \ref{Kagome}).  
For simplicity, the cut numbers are referenced against a theoretical state ${| C^{\prime} \rangle}$ in which $ Hx, Hy = 0$.  Although this reference state is unattainable by construction, it is convenient in that it makes the cut numbers directly equivalent to the winding numbers $w_x,w_y$.   Our definition of the cuts means they 
are  bound between $0 \leq Hx,Hy \leq L$ rather than the original range of [-L,L] given for the winding numbers of Rokshar and Kivelson, as discussed more below.

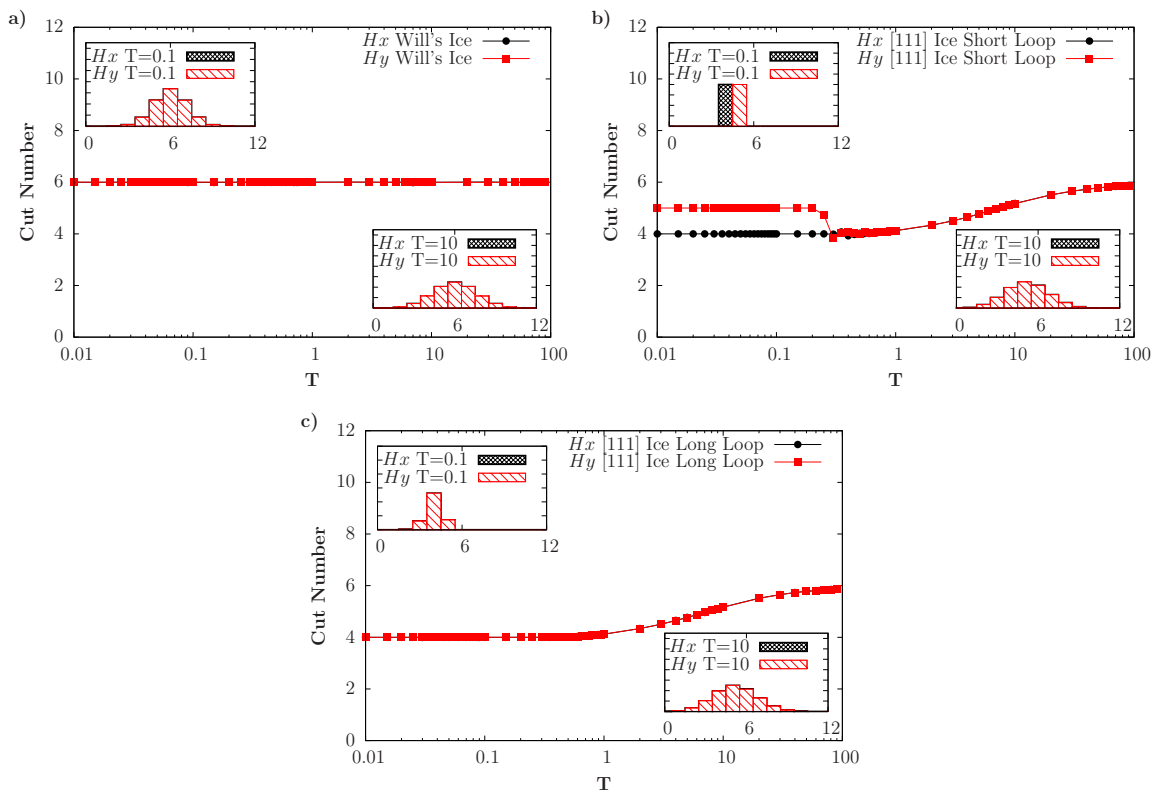
\begin{figure}[ht] 
\begin{center}
\resizebox{3.0in}{!} {\input{WI_cut.tex}}
\resizebox{3.0in}{!} {\input{SL_cut.tex}} 
\resizebox{3.0in}{!} {\input{LL_cut.tex}}  
\caption{
Topological cut numbers on an $L=12$ system for:
(a) Wills Ice ($h=0$), \\
(b) [111] Ice ($h=2$) with Short Loops,
(c) [111] Ice ($h=2$) with Long Loops.} \label{Hcuts}
\end{center}
\end{figure}

We now present simulation results for the topological cut numbers %and the pseudo- and real spin susceptibilities 
on finite-temperature Monte Carlo simulations of both [111]  ice ($h=2J$) and Wills ice ($h=0$).  
%In each figure, 
Finite-temperature annealing simulations were performed on Wills ice (which does not enter a topologically ordered phase) and on [111] ice.  In the latter case, simulations were performed with long loops, and also with short loops restricted to 
 $N_{\rm loop}^{\rm max} < L$ size. 

In Figure \ref{Hcuts} we illustrate the average cut number as a function of temperature for Wills ice and [111] ice with short and long loops.  In Wills ice the average cut numbers remain constant over the entire temperature range at a value consistent with a random and unconstrained distribution, that is with $\langle H_x \rangle /L=\langle H_y \rangle /L=1/2$ . This null result clearly illustrates the absence of topological information in the unconstrained system, with the mean cut number remaining unchanged as one passes from the high to low temperature regime. For the constrained system the cut number evolves continuously on a temperature scale between $1<T/J <10$ as the topological constraints are imposed. For the long loop simulations the mean cut numbers divided by $L$ approach $\langle H_x \rangle /L= \langle H_y \rangle /L=1/3$ and their values fluctuate around this mean. That is, the topological sectors fluctuate around a mean value consistent with broken $Z_2$ symmetry, in which exactly $1/3$ of the pseudo spins point up and the system is topologically constrained (but not topologically ordered). For the short loop simulation the behaviour is quite different: below the temperature scale on which the constraints are imposed the cut numbers freeze into fixed values, $H_x=4$ and $H_y=5$ in this case, and the system becomes topologically ordered.  Restricting to short loops therefore leads to loss of ergodicity and topological order as the constraints are imposed.

The cut number described above is a non-local quantity, measured directly by circumscribing the toroidal geometry of the simulation cell.  By definition, topological sectors require non-local measurements in order to be quantified, but such measures are notoriously inaccessible to experiment. Hence one must address the question of whether local 
measurements can be indicators of topological order in this system.  For this reason we now examine susceptibilities and their capacity to give information concerning the topological order of a constrained system.  We first define a pseudospin magnetization:
\begin{eqnarray}
\langle m \rangle = \frac{1}{N} \sum_i \sigma_i,
\label{eq:m}
\end{eqnarray}
and an associated pseudospin susceptibility,
\begin{eqnarray}
\chi_{\rho} = \frac{N}{T}[\langle m^2 \rangle - \langle m \rangle^2].
\label{eq:PS}
\end{eqnarray}
% is given in Equation \ref{eq:m} below and was used to determine the pseudospin susceptibility as shown in Equation \ref{eq:PS} for a system of size $N$ at temperature $T$.
Although, in this case, $\chi_{\rho}$ could be accessed through polarized neutron scattering on spin ice materials in the $[111]$ geometry \cite{Moessner03,Kice3} a more experimentally accessible quantity is the real spin susceptibility. The magnetization and susceptibility associated with these degrees of freedom are defined:
\begin{eqnarray}
\langle {\bf M}^\alpha \rangle = \frac{1}{N} \sum_i {\bf S^\alpha_i}  \ \ \ \ \  {\rm where}  \ \ \ \ \alpha = x,y,z;
\end{eqnarray}
\begin{eqnarray}
\chi_{r} = \frac{N}{T}[ \langle {\bf M \cdot M} \rangle - \langle {\bf M} \rangle^2].
\end{eqnarray} 
Note the $\chi_r$ can equivalently be written directly in terms of the microscopic degrees of freedom:
\begin{eqnarray}
\chi_{r} = \frac{1}{TN} \sum_{i,j}[\langle {\bf S}_i \cdot {\bf S}_j \rangle - \langle {\bf S}_i\rangle \langle {\bf S}_j \rangle ]
\end{eqnarray}

 \begin{figure}[] 
\begin{center}
\resizebox{3.0in}{!} {\input{Pseudo_Sus.tex}}
\resizebox{3.0in}{!} {\input{Real_Sus.tex}}
\caption{
(a) Pseudospin susceptibility, and (b) Real spin susceptibility.  
Each case is for an $L=12$ system. Results for unconstrained Wills Ice are in black, [111] long loop spin ice in red, and [111] short loop spin ice in blue.
}
\label{susc}
\end{center}
\end{figure}
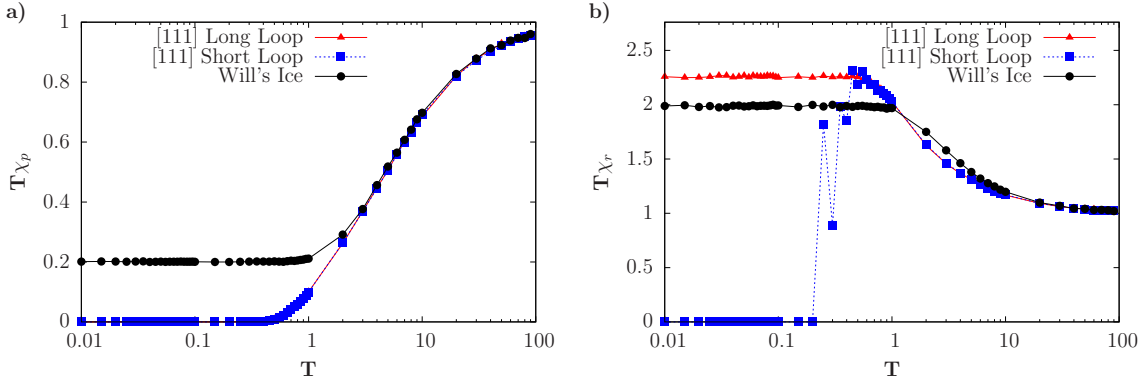

We simulate both the pseudo and real spin susceptibilities, for both Wills and [111] ice, using the same procedures as for the cut numbers, described above.
Pseudospin susceptibilities are illustrated in Figure \ref{susc}(a).  Here, one sees that $T\chi_{\rho}$ reaches a limiting value for both Wills and [111] kagome ice for $T < 1$, which corresponds to the temperature where the system settles into the disordered groundstate manifold.  The low-temperature value of $T\chi_{\rho}$ is zero for  [111] kagome ice, indicating that fluctuations in pseudospin magnetization are frozen out.  This result is consistent with the topological constraints being enforced everywhere on a temperature scale less than $h$, so that once the number of triangles not satisfying the $uud$ constraint becomes exponentially small, the fluctuations in $m$ fall exponentially to zero. This is true for both the long and short loop algorithm and so is a measure of the applied topological constraints, but is not a measure of topological order. This result should be contrasted with that for Wills ice whose pseudo spin susceptibility continues to fluctuate at low temperature, illustrating that it is exploring the full low temperature phase space of $upp$ and $uud$ states.

Figure \ref{susc}(b) illustrates $T\chi_r$ as a function of temperature, which also shows different behaviour as  $T \rightarrow 0$ for our different models.  Wills ice and the long loop simulation for $[111]$ both approach finite values as $T$ approaches zero.  There is, however a 
most striking difference between long and short loop simulations for  [111] ice.  For short loops, $T\chi_r$ drops abruptly to zero for $T\approx 0.5$. From Figure \ref{Hcuts} one can see that this drop corresponds to the onset of topological order. As the sectors can only change by introducing a loop that spans the periodic boundaries, when such loops are suppressed the sector becomes frozen. This sector freezing manifests itself in the magnetic fluctuations, as short closed loops inside the system carry no magnetization and contribute zero to $T\chi_r$, so that it falls abruptly to zero as the concentration of topological defects disappears. System-spanning long loops do however carry magnetization, and the finite value for $T\chi_r$ at low temperature is consistent with the fluctuating topological sector at low temperature. Restricting loop length therefore leads to a dynamical phase transition to a state with long range topological order and this evolution is apparent in the real spin susceptibility.

\begin{figure}%[h!] 
\begin{center}
\resizebox{3.2in}{!} {\input{RS_long.tex}}
\caption{Measure of the real susceptibility in [111] ice for system sizes from $L=32$ to $L=3$ at a constant temperature of $T=0.01$.  The temperature multiplied by the real susceptibility is plotted against the inverse linear system size.}
\label{RS_long}
\end{center}
\end{figure}
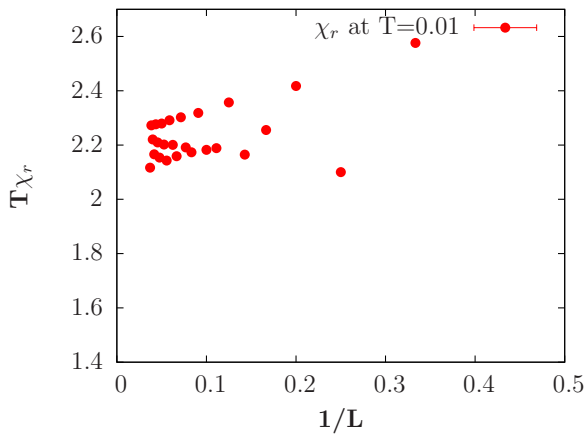

We can estimate the limiting value $T\chi_r$ as $T \rightarrow 0$ by approximating the kagome lattice by a Hushimi tree of three-fold coordinated corner sharing triangles (see the Appendix) and summing the combinatorial series corresponding to the ground state manifold of states \cite{Ludovic}. Such tree calculations have proved to be extremely accurate for the nearest neighbour spin ice model \cite{Ludovic} despite the fact that the closed nearest neighbour pathways appearing on the d-dimensional lattice are neglected. As in the case of nearest neighbour spin ice and the six vertex model on a square lattice \cite{Louis-Paul}, the tree calculation yields $T\chi_r(L = \infty) = 2$, while finite-size scaling (Figure \ref{RS_long}) at $T=0.1$ for [111] ice with long loops indicates that $T\chi_r(L = \infty) \approx 2.2$. For the constrained system, the closed pathways seem to have a considerable influence on the low temperature susceptibility. Interestingly, this is not the case for Wills ice: here, the tree calculation also yields the modified Curie law, $T\chi_r(L = \infty) = 2$ and simulation yields the same result within numerical accuracy.  This agreement is consistent with the absence of non-local correlations in the unconstrained system.

\begin{figure}%[h!] 
\begin{center}
\resizebox{3.2in}{!} {\input{RS_subset.tex}}
\caption{Measure of the real susceptibility in [111] ice for for a varying subset $\ell$ of a $L=32$ lattice. This data is taken from several simulations which were identical except for their loop algorithm, which was limited to produce loops of a specific size in each run.}
\label{RS_sub}
\end{center}
\end{figure}
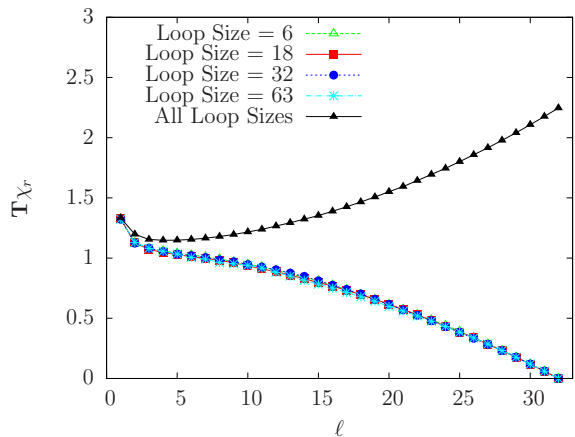

Finally, we can examine the effect of restricted loop size on the real susceptibility of a subset of the topologically ordered system.  Since we have seen that restricted local dynamics are a critical ingredient for the non-ergodicity between topological sectors, we can examine the effect of loop size on subset regions of varying size.  In doing these simulations we are able to test the effect of different boundary conditions, as small windows in a simulation of much greater size, approach the situation of open boundaries. Performing simulations on a $L=32$ system, we measure $\chi_r$ on a subset of spins with size $\ell \times \ell \times 3$ at $T=0.01$.  From Figure \ref{susc}  for loops restricted to any size  smaller than $L$, $T\chi_r =0$ and the system will be in a topologically ordered phase.  In Figure \ref{RS_sub} we show susceptibility results for loop size less than $L$, as a function of linear subregion size $\ell$.  Interestingly, $T\chi_r \neq 0$ for all $\ell < L$, even if $\ell$ is larger than the maximum allowed loop size.  Magnetic fluctuations are allowed within the widow of size $\ell$, even though the total topological sector is frozen, as short loops spanning the edge of the window carry magnetization. 
As $\ell$ gets larger than the maximum loop size, all magnetic fluctuations result from edge loops, hence $T\chi_r$ versus $\ell$ scales linearly to zero as $\ell \rightarrow L$. As $\ell$ increases the results become independent of the maximum loop length, as long as this is less than $L$. This suggests that the major contribution to this measure comes from short loops, which is rather surprising given that the loop length distribution is expected to be power law up to the imposed cut off. In contrast, if the loop lengths are unrestricted the susceptibility of the window increases monotonically towards the fluctuating sector susceptibility as $\ell$ approaches $L$.
These results illustrate the non-local nature of the topological information encoded into the system. The window susceptibility for the two cases is the same if $\ell \ll L$, but their relative difference becomes of order one as $\ell$ increases and the effect of the boundaries are felt over macroscopic length scales.

\section{Discussion}

In this paper we have employed Monte Carlo simulations using local and non-local loop updates to characterize the classical topological fluctuations  occurring in the ground state of the constrained kagome ice model, realized by placing an external magnetic field along the [111] direction of the nearest neighbour spin ice model. We characterize the topological sector in the groundstate using a dual-lattice dimer representation: topological numbers are defined by taking a ``cut'' across dimers directed from the A to B sublattice of the dual (honeycomb) lattice.  Defined in this way, we show that Monte Carlo simulations using local dynamics (including size-restricted loop updates) loose ergodicity between topological sectors as the system settles into the groundstate manifold of states, leading to a topologically ordered state.  Unrestricted loop updates, which allow loops to wind around the periodic boundary of the system, allow the system to fluctuate between topological sectors at all temperatures.

Defining a susceptibility $\chi_r$ using the 2D projection of the real spin ice moments, we show that the onset of topological order has signatures in $T\chi_r$.  Specifically, when the system is not allowed to fluctuate between topological sectors in the groundstate (i.e. in the presence of local updates), $T\chi_r = 0$.  If ergodicity between topological sectors is restored by allowing loop Monte Carlo moves of arbitrary length, $T\chi_r$ remains finite down to $T \rightarrow 0$. As the system settles into the ground state manifold, the susceptibility crosses over from Curie's law, $\chi =1/T$ to a modified Curie law reflecting the hidden non-local correlations, which we estimate numerically to be   $T \chi_r \approx 2.2$. This is slightly above the value estimated from a calculation of triangular spin units on a Husami tree, $T \chi_r = 2$, showing the importance of closed pathways of spins in constrained kagome ice.

Throughout the paper we have compared and contrasted the physics of  constrained ($[111]$) kagome ice with that of unconstrained (Wills) kagome ice.  We have illustrated how the constraints lead to the emergence of a hidden $U(1)$ gauge field and to a highly correlated liquid state.   If we take the full, unconstrained phase space of Wills ice, there is no emergent gauge field and the excitations out of the ground state manifold do not carry topological charge. In analogy with spin ice, monopole physics has previously been discussed in the context of artificial magnetic arrays with nano-magnets arranged on a honeycomb lattice \cite{kag-mon-1,kag-mon-2}.  There, the phase space of orientations for the nano-elements is the same as that for the real spins in the kagome ice models discussed in this paper. However, in order to realize a gas of monopole quasi particles in two dimensions, one must have both an underlying emergent gauge field and effective Coulomb interactions between the topological excitations out of the constrained manifold of low energy states. The first criterion can therefore only be satisfied if the $Z_2$ symmetry is broken and topological constraints are imposed. Although there is some evidence that dipole-dipole interactions between elements could spontaneously break this symmetry \cite{charge-order}, it does not appear to be the case in experimental presented so far, with the result that the monopole description can only be valid in the low density limit, with excitations above a specific ordered array of nano-elements.

In conclusion, we have shown that a precise definition of the topological number or sector must involve a non-local cut measure which circumscribes the toroidal geometry of the simulation cell. However, by measuring the susceptibility for spins in windows contained within a bigger system, we have demonstrated that $T\chi_r$ can be used to determine whether or not a system is ergodic between topological sectors.  By increasing the window size, or measuring spin fluctuations on increasing scales, on finds scaling towards two well-defined limits, depending on whether topological sector fluctuations occur or not. Susceptibility measurements such as this may prove to be a valuable probe of topological order on classical systems such as spin ice in the future.

\section{Acknowledgments}

We thank A. Harman-Clark,  M. Hastings, L. P. Henry,  L. Jaubert and T. Roscilde for enlightening discussions.  R.G.M. thanks the Ecole Normale Superieure de Lyon for hospitality during visits and the French Embassy in Canada for a Canada-France travel grant.  
Simulations were performed using the computing facilities of SHARCNET.
This work was supported by NSERC of Canada (A.J.M. and R.G.M.) and the National Science Foundation under grant NSF PHY05-51164 (R.G.M.)

\section*{References}
%\begin{thebibliography}{99}
%
%\end{thebibliography}

\bibliographystyle{unsrt}
\bibliography{Biblio}

\section{Appendix: Limiting Value of the Real Susceptibility}

Adapting the argument laid out in the thesis of Jaubert \cite{Ludovic} for pyrochlore spin ice, it is possible to construct limiting values for the real susceptibility within the kagome plane. Assuming that the system approximates a grand canonical ensemble of infinite size such as a Hushimi tree, rather than a lattice of fixed size with periodic boundary conditions, one can proceed by expanding the sums in the expression for the real susceptibility as below.

\begin{eqnarray}
\chi_{r} =  \frac{1}{T}\left[ { \frac{1}{N} \sum_i  \langle {\bf S}_i^2 \rangle + \sum_{i \not= 0} \langle {\bf S}_i \cdot {\bf S}_0 \rangle }\right]
\end{eqnarray}

Since $\langle {\bf S}_i \rangle = 0$ and $\langle {\bf S}_i^2 \rangle = 1  \ \forall \  i$:
 \begin{eqnarray}
\chi_{r} =  \frac{1}{T} \left[{ 1 + \sum_{i \not= 0} \langle {\bf S}_i \cdot {\bf S}_0 \rangle }\right]
\label{eq:sum}
\end{eqnarray}

The Curie Law $\chi_r ={1}/{T}$, is recovered in the paramagnet phase, as when the constraints of the ice rules are broken there is no correlation between nearest neighbors. As increased thermal excitations cause a breakdown of the ice rules the Curie law thus dictates the limiting behavior for high temperature. To derive the limiting behavior for low temperatures we rewrite the sum in Equation \ref{eq:sum} as an infinite sum over all nearest neighbors.

\begin{eqnarray}
\sum_{i \not= 0} \langle {\bf S}_i \cdot {\bf S}_0 \rangle = \sum_{n = 1}^{\infty} \sum_{i \in n^{th}NN} \langle {\bf S}_i \cdot {\bf S}_0 \rangle
\label{eq:NNsum}
\end{eqnarray}

\begin{figure}%[h!] 
\begin{center}
\includegraphics[width=3.5in]{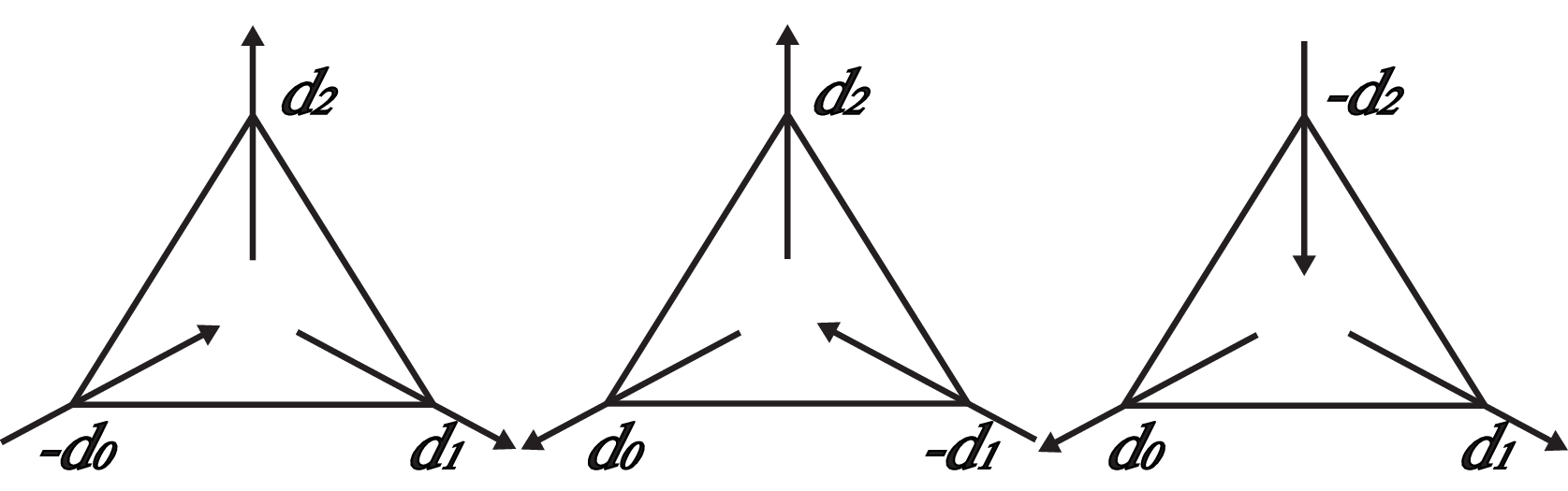}
\caption{States that obey the ice rules within the kagome plane, with accompanying unit vectors $\hat{d}_i$.}
\label{States}
\end{center}
\end{figure}

Choosing an arbitrary spin in the plane to be ${\bf S}_0$ the dot product can be computed between ${\bf S}_0$ and its four nearest neighbors ${\bf S}_i$. Treating the out of plane spin as fixed by the [111] external field, there are three potential spin configurations (see Figure \ref{States}) that satisfy the ice rules in the kagome plane, with the following expectation values. 

 \label{states}

 \begin{eqnarray}
\sum_{i \not= 0} \langle  {\bf S}_i \cdot {\bf S}_0 \rangle = \frac{1}{2} ( \frac{1}{2} - \frac{1}{2} ) = 0\\
\sum_{i \not= 0} \langle {\bf S}_i \cdot {\bf S}_0 \rangle = \frac{1}{2} ( -\frac{1}{2} + \frac{1}{2} ) = 0\\
\sum_{i \not= 0}  \langle {\bf S}_i \cdot {\bf S}_0 \rangle = \frac{1}{2} ( \frac{1}{2} + \frac{1}{2} ) = \frac{1}{2}\\
 \end{eqnarray}
 
The average value of all three configurations is:
\begin{eqnarray}
\sum_{i \not= 0} \langle {\bf S}_i \cdot {\bf S}_0\rangle_{all \ configs.} = \frac{1}{3}(0+0+\frac{1}{2}) = \frac{1}{6}
 \end{eqnarray}
 
Since there are four nearest neighbors for every spin the number of nth nearest neighbors is given by $4(2^{n-1}) = 2^{n+1}$. Taking the sum to infinity gives:

\begin{eqnarray}
\sum_{n=1}^{\infty} g_n = \sum_{n=1}^{\infty}   2^{n+1} (\frac{1}{6})^n = \sum_{n=1}^{\infty} \frac{2}{3^n} = 1
 \end{eqnarray}
 
 Leading to a limiting value of the real susceptibility identical to that found for the 3-D Hushimi tree:
 \begin{eqnarray}
 \chi_r(T \rightarrow 0) \sim \frac{2}{T}.
  \end{eqnarray}
  
This result also follows for Wills Ice, and can be derived through the same counting argument by ignoring the restrictions imposed by the ice rules.  This leads to three more states of minimum energy, specifically the two-in/one-out states, contributing to the nearest neighbor inner product sum.  However, the three new states have the exact same averaged sum as the spin ice accessible states. Thus the sum over all possible configurations and accompanying real susceptibility are the same as in [111] ice.
  
In the Monte Carlo simulation, these results were found to accurately reflect the limiting value of the real susceptibility of Wills Ice, even for small system size.  However, once the external magnetic field was added to the Hamiltonian the topological constraints imposed on the boundaries via the ice rules prevented the system from reaching the theoretical $T \rightarrow 0$ limit for [111] ice, regardless of system size.

\end{document}

%% file: WI_cut.tex
% GNUPLOT: LaTeX picture with Postscript
\begingroup
  \makeatletter
  \providecommand\color[2][]{%
    \GenericError{(gnuplot) \space\space\space\@spaces}{%
      Package color not loaded in conjunction with
      terminal option `colourtext'%
    }{See the gnuplot documentation for explanation.%
    }{Either use 'blacktext' in gnuplot or load the package
      color.sty in LaTeX.}%
    \renewcommand\color[2][]{}%
  }%
  \providecommand\includegraphics[2][]{%
    \GenericError{(gnuplot) \space\space\space\@spaces}{%
      Package graphicx or graphics not loaded%
    }{See the gnuplot documentation for explanation.%
    }{The gnuplot epslatex terminal needs graphicx.sty or graphics.sty.}%
    \renewcommand\includegraphics[2][]{}%
  }%
  \providecommand\rotatebox[2]{#2}%
  \@ifundefined{ifGPcolor}{%
    \newif\ifGPcolor
    \GPcolortrue
  }{}%
  \@ifundefined{ifGPblacktext}{%
    \newif\ifGPblacktext
    \GPblacktexttrue
  }{}%
  % define a \g@addto@macro without @ in the name:
  \let\gplgaddtomacro\g@addto@macro
  % define empty templates for all commands taking text:
  \gdef\gplbacktext{}%
  \gdef\gplfronttext{}%
  \makeatother
  \ifGPblacktext
    % no textcolor at all
    \def\colorrgb#1{}%
    \def\colorgray#1{}%
  \else
    % gray or color?
    \ifGPcolor
      \def\colorrgb#1{\color[rgb]{#1}}%
      \def\colorgray#1{\color[gray]{#1}}%
      \expandafter\def\csname LTw\endcsname{\color{white}}%
      \expandafter\def\csname LTb\endcsname{\color{black}}%
      \expandafter\def\csname LTa\endcsname{\color{black}}%
      \expandafter\def\csname LT0\endcsname{\color[rgb]{1,0,0}}%
      \expandafter\def\csname LT1\endcsname{\color[rgb]{0,1,0}}%
      \expandafter\def\csname LT2\endcsname{\color[rgb]{0,0,1}}%
      \expandafter\def\csname LT3\endcsname{\color[rgb]{1,0,1}}%
      \expandafter\def\csname LT4\endcsname{\color[rgb]{0,1,1}}%
      \expandafter\def\csname LT5\endcsname{\color[rgb]{1,1,0}}%
      \expandafter\def\csname LT6\endcsname{\color[rgb]{0,0,0}}%
      \expandafter\def\csname LT7\endcsname{\color[rgb]{1,0.3,0}}%
      \expandafter\def\csname LT8\endcsname{\color[rgb]{0.5,0.5,0.5}}%
    \else
      % gray
      \def\colorrgb#1{\color{black}}%
      \def\colorgray#1{\color[gray]{#1}}%
      \expandafter\def\csname LTw\endcsname{\color{white}}%
      \expandafter\def\csname LTb\endcsname{\color{black}}%
      \expandafter\def\csname LTa\endcsname{\color{black}}%
      \expandafter\def\csname LT0\endcsname{\color{black}}%
      \expandafter\def\csname LT1\endcsname{\color{black}}%
      \expandafter\def\csname LT2\endcsname{\color{black}}%
      \expandafter\def\csname LT3\endcsname{\color{black}}%
      \expandafter\def\csname LT4\endcsname{\color{black}}%
      \expandafter\def\csname LT5\endcsname{\color{black}}%
      \expandafter\def\csname LT6\endcsname{\color{black}}%
      \expandafter\def\csname LT7\endcsname{\color{black}}%
      \expandafter\def\csname LT8\endcsname{\color{black}}%
    \fi
  \fi
  \setlength{\unitlength}{0.0500bp}%
  \begin{picture}(7200.00,5040.00)%
    \gplgaddtomacro\gplbacktext{%
      \csname LTb\endcsname%
      \put(748,704){\makebox(0,0)[r]{\strut{} 0}}%
      \put(748,1353){\makebox(0,0)[r]{\strut{} 2}}%
      \put(748,2002){\makebox(0,0)[r]{\strut{} 4}}%
      \put(748,2652){\makebox(0,0)[r]{\strut{} 6}}%
      \put(748,3301){\makebox(0,0)[r]{\strut{} 8}}%
      \put(748,3950){\makebox(0,0)[r]{\strut{} 10}}%
      \put(748,4599){\makebox(0,0)[r]{\strut{} 12}}%
      \put(880,484){\makebox(0,0){\strut{} 0.01}}%
      \put(2377,484){\makebox(0,0){\strut{} 0.1}}%
      \put(3875,484){\makebox(0,0){\strut{} 1}}%
      \put(5372,484){\makebox(0,0){\strut{} 10}}%
      \put(6869,484){\makebox(0,0){\strut{} 100}}%
      \put(308,2651){\rotatebox{-270}{\makebox(0,0){\strut{}\textbf{Cut Number}}}}%
      \put(3874,154){\makebox(0,0){\strut{}\textbf{T}}}%
      \put(178,4709){\makebox(0,0){\strut{}\textbf{a)}}}%
    }%
    \gplgaddtomacro\gplfronttext{%
      \put(5882,4426){\makebox(0,0)[r]{\strut{}$Hx$ Will's Ice}}%
      \csname LTb\endcsname%
      \put(5882,4206){\makebox(0,0)[r]{\strut{}$Hy$ Will's Ice}}%
    }%
    \gplgaddtomacro\gplbacktext{%
      \csname LTb\endcsname%
      \put(898,3355){\makebox(0,0)[r]{\strut{}}}%
      \put(898,3495){\makebox(0,0)[r]{\strut{}}}%
      \put(898,3635){\makebox(0,0)[r]{\strut{}}}%
      \put(898,3775){\makebox(0,0)[r]{\strut{}}}%
      \put(898,3915){\makebox(0,0)[r]{\strut{}}}%
      \put(898,4055){\makebox(0,0)[r]{\strut{}}}%
      \put(898,4195){\makebox(0,0)[r]{\strut{}}}%
      \put(898,4335){\makebox(0,0)[r]{\strut{}}}%
      \put(1030,3135){\makebox(0,0){\strut{} 0}}%
      \put(2092,3135){\makebox(0,0){\strut{} 6}}%
      \put(3154,3135){\makebox(0,0){\strut{} 12}}%
    }%
    \gplgaddtomacro\gplfronttext{%
      \put(2167,4232){\makebox(0,0)[r]{\strut{}$Hx$ T=0.1}}%
      \csname LTb\endcsname%
      \put(2167,4012){\makebox(0,0)[r]{\strut{}$Hy$ T=0.1}}%
    }%
    \gplgaddtomacro\gplbacktext{%
      \csname LTb\endcsname%
      \put(4506,1070){\makebox(0,0)[r]{\strut{}}}%
      \put(4506,1201){\makebox(0,0)[r]{\strut{}}}%
      \put(4506,1332){\makebox(0,0)[r]{\strut{}}}%
      \put(4506,1464){\makebox(0,0)[r]{\strut{}}}%
      \put(4506,1595){\makebox(0,0)[r]{\strut{}}}%
      \put(4506,1726){\makebox(0,0)[r]{\strut{}}}%
      \put(4506,1857){\makebox(0,0)[r]{\strut{}}}%
      \put(4506,1988){\makebox(0,0)[r]{\strut{}}}%
      \put(4638,850){\makebox(0,0){\strut{} 0}}%
      \put(5664,850){\makebox(0,0){\strut{} 6}}%
      \put(6689,850){\makebox(0,0){\strut{} 12}}%
    }%
    \gplgaddtomacro\gplfronttext{%
      \put(5702,1881){\makebox(0,0)[r]{\strut{}$Hx$ T=10}}%
      \csname LTb\endcsname%
      \put(5702,1661){\makebox(0,0)[r]{\strut{}$Hy$ T=10}}%
    }%
    \gplbacktext
    \put(0,0){\includegraphics{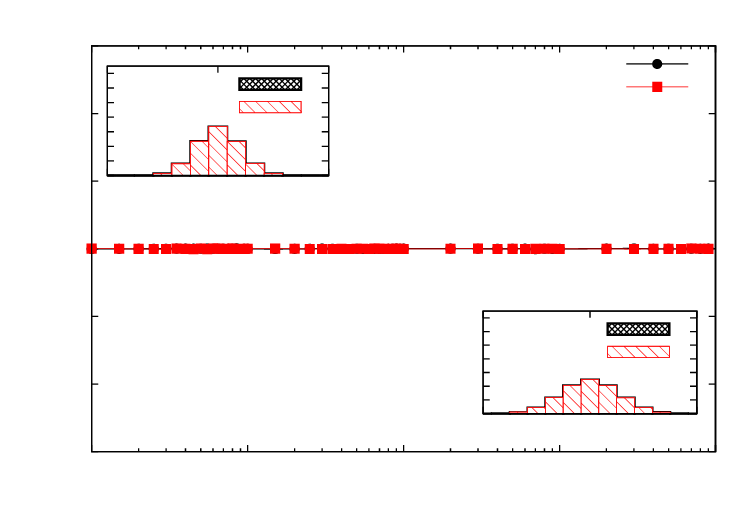}}%
    \gplfronttext
  \end{picture}%
\endgroup

%% file: SL_cut.tex
% GNUPLOT: LaTeX picture with Postscript
\begingroup
  \makeatletter
  \providecommand\color[2][]{%
    \GenericError{(gnuplot) \space\space\space\@spaces}{%
      Package color not loaded in conjunction with
      terminal option `colourtext'%
    }{See the gnuplot documentation for explanation.%
    }{Either use 'blacktext' in gnuplot or load the package
      color.sty in LaTeX.}%
    \renewcommand\color[2][]{}%
  }%
  \providecommand\includegraphics[2][]{%
    \GenericError{(gnuplot) \space\space\space\@spaces}{%
      Package graphicx or graphics not loaded%
    }{See the gnuplot documentation for explanation.%
    }{The gnuplot epslatex terminal needs graphicx.sty or graphics.sty.}%
    \renewcommand\includegraphics[2][]{}%
  }%
  \providecommand\rotatebox[2]{#2}%
  \@ifundefined{ifGPcolor}{%
    \newif\ifGPcolor
    \GPcolortrue
  }{}%
  \@ifundefined{ifGPblacktext}{%
    \newif\ifGPblacktext
    \GPblacktexttrue
  }{}%
  % define a \g@addto@macro without @ in the name:
  \let\gplgaddtomacro\g@addto@macro
  % define empty templates for all commands taking text:
  \gdef\gplbacktext{}%
  \gdef\gplfronttext{}%
  \makeatother
  \ifGPblacktext
    % no textcolor at all
    \def\colorrgb#1{}%
    \def\colorgray#1{}%
  \else
    % gray or color?
    \ifGPcolor
      \def\colorrgb#1{\color[rgb]{#1}}%
      \def\colorgray#1{\color[gray]{#1}}%
      \expandafter\def\csname LTw\endcsname{\color{white}}%
      \expandafter\def\csname LTb\endcsname{\color{black}}%
      \expandafter\def\csname LTa\endcsname{\color{black}}%
      \expandafter\def\csname LT0\endcsname{\color[rgb]{1,0,0}}%
      \expandafter\def\csname LT1\endcsname{\color[rgb]{0,1,0}}%
      \expandafter\def\csname LT2\endcsname{\color[rgb]{0,0,1}}%
      \expandafter\def\csname LT3\endcsname{\color[rgb]{1,0,1}}%
      \expandafter\def\csname LT4\endcsname{\color[rgb]{0,1,1}}%
      \expandafter\def\csname LT5\endcsname{\color[rgb]{1,1,0}}%
      \expandafter\def\csname LT6\endcsname{\color[rgb]{0,0,0}}%
      \expandafter\def\csname LT7\endcsname{\color[rgb]{1,0.3,0}}%
      \expandafter\def\csname LT8\endcsname{\color[rgb]{0.5,0.5,0.5}}%
    \else
      % gray
      \def\colorrgb#1{\color{black}}%
      \def\colorgray#1{\color[gray]{#1}}%
      \expandafter\def\csname LTw\endcsname{\color{white}}%
      \expandafter\def\csname LTb\endcsname{\color{black}}%
      \expandafter\def\csname LTa\endcsname{\color{black}}%
      \expandafter\def\csname LT0\endcsname{\color{black}}%
      \expandafter\def\csname LT1\endcsname{\color{black}}%
      \expandafter\def\csname LT2\endcsname{\color{black}}%
      \expandafter\def\csname LT3\endcsname{\color{black}}%
      \expandafter\def\csname LT4\endcsname{\color{black}}%
      \expandafter\def\csname LT5\endcsname{\color{black}}%
      \expandafter\def\csname LT6\endcsname{\color{black}}%
      \expandafter\def\csname LT7\endcsname{\color{black}}%
      \expandafter\def\csname LT8\endcsname{\color{black}}%
    \fi
  \fi
  \setlength{\unitlength}{0.0500bp}%
  \begin{picture}(7200.00,5040.00)%
    \gplgaddtomacro\gplbacktext{%
      \csname LTb\endcsname%
      \put(748,704){\makebox(0,0)[r]{\strut{} 0}}%
      \put(748,1353){\makebox(0,0)[r]{\strut{} 2}}%
      \put(748,2002){\makebox(0,0)[r]{\strut{} 4}}%
      \put(748,2652){\makebox(0,0)[r]{\strut{} 6}}%
      \put(748,3301){\makebox(0,0)[r]{\strut{} 8}}%
      \put(748,3950){\makebox(0,0)[r]{\strut{} 10}}%
      \put(748,4599){\makebox(0,0)[r]{\strut{} 12}}%
      \put(880,484){\makebox(0,0){\strut{} 0.01}}%
      \put(2377,484){\makebox(0,0){\strut{} 0.1}}%
      \put(3875,484){\makebox(0,0){\strut{} 1}}%
      \put(5372,484){\makebox(0,0){\strut{} 10}}%
      \put(6869,484){\makebox(0,0){\strut{} 100}}%
      \put(308,2651){\rotatebox{-270}{\makebox(0,0){\strut{}\textbf{Cut Number}}}}%
      \put(3874,154){\makebox(0,0){\strut{}\textbf{T}}}%
      \put(178,4709){\makebox(0,0){\strut{}\textbf{b)}}}%
    }%
    \gplgaddtomacro\gplfronttext{%
      \put(5882,4426){\makebox(0,0)[r]{\strut{}$Hx$ [111] Ice Short Loop}}%
      \csname LTb\endcsname%
      \put(5882,4206){\makebox(0,0)[r]{\strut{}$Hy$ [111] Ice Short Loop}}%
    }%
    \gplgaddtomacro\gplbacktext{%
      \csname LTb\endcsname%
      \put(898,3355){\makebox(0,0)[r]{\strut{}}}%
      \put(898,3486){\makebox(0,0)[r]{\strut{}}}%
      \put(898,3618){\makebox(0,0)[r]{\strut{}}}%
      \put(898,3749){\makebox(0,0)[r]{\strut{}}}%
      \put(898,3880){\makebox(0,0)[r]{\strut{}}}%
      \put(898,4011){\makebox(0,0)[r]{\strut{}}}%
      \put(898,4143){\makebox(0,0)[r]{\strut{}}}%
      \put(898,4274){\makebox(0,0)[r]{\strut{}}}%
      \put(898,4405){\makebox(0,0)[r]{\strut{}}}%
      \put(1030,3135){\makebox(0,0){\strut{} 0}}%
      \put(2092,3135){\makebox(0,0){\strut{} 6}}%
      \put(3154,3135){\makebox(0,0){\strut{} 12}}%
    }%
    \gplgaddtomacro\gplfronttext{%
      \put(2167,4232){\makebox(0,0)[r]{\strut{}$Hx$ T=0.1}}%
      \csname LTb\endcsname%
      \put(2167,4012){\makebox(0,0)[r]{\strut{}$Hy$ T=0.1}}%
    }%
    \gplgaddtomacro\gplbacktext{%
      \csname LTb\endcsname%
      \put(4506,1070){\makebox(0,0)[r]{\strut{}}}%
      \put(4506,1201){\makebox(0,0)[r]{\strut{}}}%
      \put(4506,1332){\makebox(0,0)[r]{\strut{}}}%
      \put(4506,1464){\makebox(0,0)[r]{\strut{}}}%
      \put(4506,1595){\makebox(0,0)[r]{\strut{}}}%
      \put(4506,1726){\makebox(0,0)[r]{\strut{}}}%
      \put(4506,1857){\makebox(0,0)[r]{\strut{}}}%
      \put(4506,1988){\makebox(0,0)[r]{\strut{}}}%
      \put(4638,850){\makebox(0,0){\strut{} 0}}%
      \put(5664,850){\makebox(0,0){\strut{} 6}}%
      \put(6689,850){\makebox(0,0){\strut{} 12}}%
    }%
    \gplgaddtomacro\gplfronttext{%
      \put(5702,1881){\makebox(0,0)[r]{\strut{}$Hx$ T=10}}%
      \csname LTb\endcsname%
      \put(5702,1661){\makebox(0,0)[r]{\strut{}$Hy$ T=10}}%
    }%
    \gplbacktext
    \put(0,0){\includegraphics{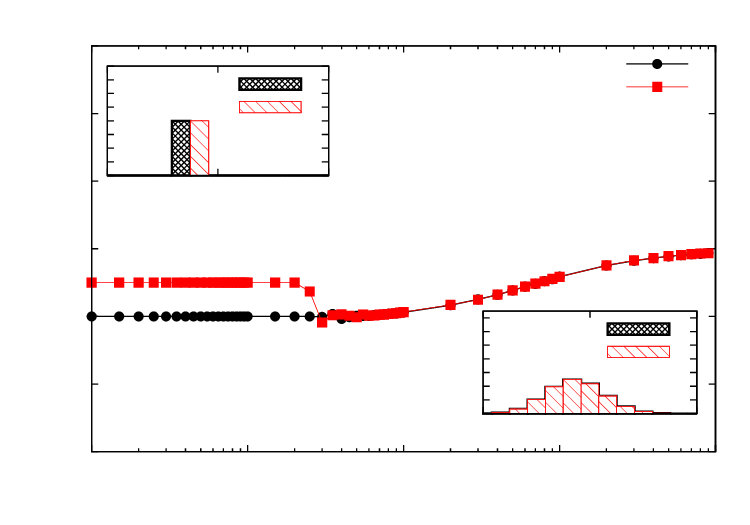}}%
    \gplfronttext
  \end{picture}%
\endgroup

%% file: LL_cut.tex
% GNUPLOT: LaTeX picture with Postscript
\begingroup
  \makeatletter
  \providecommand\color[2][]{%
    \GenericError{(gnuplot) \space\space\space\@spaces}{%
      Package color not loaded in conjunction with
      terminal option `colourtext'%
    }{See the gnuplot documentation for explanation.%
    }{Either use 'blacktext' in gnuplot or load the package
      color.sty in LaTeX.}%
    \renewcommand\color[2][]{}%
  }%
  \providecommand\includegraphics[2][]{%
    \GenericError{(gnuplot) \space\space\space\@spaces}{%
      Package graphicx or graphics not loaded%
    }{See the gnuplot documentation for explanation.%
    }{The gnuplot epslatex terminal needs graphicx.sty or graphics.sty.}%
    \renewcommand\includegraphics[2][]{}%
  }%
  \providecommand\rotatebox[2]{#2}%
  \@ifundefined{ifGPcolor}{%
    \newif\ifGPcolor
    \GPcolortrue
  }{}%
  \@ifundefined{ifGPblacktext}{%
    \newif\ifGPblacktext
    \GPblacktexttrue
  }{}%
  % define a \g@addto@macro without @ in the name:
  \let\gplgaddtomacro\g@addto@macro
  % define empty templates for all commands taking text:
  \gdef\gplbacktext{}%
  \gdef\gplfronttext{}%
  \makeatother
  \ifGPblacktext
    % no textcolor at all
    \def\colorrgb#1{}%
    \def\colorgray#1{}%
  \else
    % gray or color?
    \ifGPcolor
      \def\colorrgb#1{\color[rgb]{#1}}%
      \def\colorgray#1{\color[gray]{#1}}%
      \expandafter\def\csname LTw\endcsname{\color{white}}%
      \expandafter\def\csname LTb\endcsname{\color{black}}%
      \expandafter\def\csname LTa\endcsname{\color{black}}%
      \expandafter\def\csname LT0\endcsname{\color[rgb]{1,0,0}}%
      \expandafter\def\csname LT1\endcsname{\color[rgb]{0,1,0}}%
      \expandafter\def\csname LT2\endcsname{\color[rgb]{0,0,1}}%
      \expandafter\def\csname LT3\endcsname{\color[rgb]{1,0,1}}%
      \expandafter\def\csname LT4\endcsname{\color[rgb]{0,1,1}}%
      \expandafter\def\csname LT5\endcsname{\color[rgb]{1,1,0}}%
      \expandafter\def\csname LT6\endcsname{\color[rgb]{0,0,0}}%
      \expandafter\def\csname LT7\endcsname{\color[rgb]{1,0.3,0}}%
      \expandafter\def\csname LT8\endcsname{\color[rgb]{0.5,0.5,0.5}}%
    \else
      % gray
      \def\colorrgb#1{\color{black}}%
      \def\colorgray#1{\color[gray]{#1}}%
      \expandafter\def\csname LTw\endcsname{\color{white}}%
      \expandafter\def\csname LTb\endcsname{\color{black}}%
      \expandafter\def\csname LTa\endcsname{\color{black}}%
      \expandafter\def\csname LT0\endcsname{\color{black}}%
      \expandafter\def\csname LT1\endcsname{\color{black}}%
      \expandafter\def\csname LT2\endcsname{\color{black}}%
      \expandafter\def\csname LT3\endcsname{\color{black}}%
      \expandafter\def\csname LT4\endcsname{\color{black}}%
      \expandafter\def\csname LT5\endcsname{\color{black}}%
      \expandafter\def\csname LT6\endcsname{\color{black}}%
      \expandafter\def\csname LT7\endcsname{\color{black}}%
      \expandafter\def\csname LT8\endcsname{\color{black}}%
    \fi
  \fi
  \setlength{\unitlength}{0.0500bp}%
  \begin{picture}(7200.00,5040.00)%
    \gplgaddtomacro\gplbacktext{%
      \csname LTb\endcsname%
      \put(748,704){\makebox(0,0)[r]{\strut{} 0}}%
      \put(748,1353){\makebox(0,0)[r]{\strut{} 2}}%
      \put(748,2002){\makebox(0,0)[r]{\strut{} 4}}%
      \put(748,2652){\makebox(0,0)[r]{\strut{} 6}}%
      \put(748,3301){\makebox(0,0)[r]{\strut{} 8}}%
      \put(748,3950){\makebox(0,0)[r]{\strut{} 10}}%
      \put(748,4599){\makebox(0,0)[r]{\strut{} 12}}%
      \put(880,484){\makebox(0,0){\strut{} 0.01}}%
      \put(2377,484){\makebox(0,0){\strut{} 0.1}}%
      \put(3875,484){\makebox(0,0){\strut{} 1}}%
      \put(5372,484){\makebox(0,0){\strut{} 10}}%
      \put(6869,484){\makebox(0,0){\strut{} 100}}%
      \put(308,2651){\rotatebox{-270}{\makebox(0,0){\strut{}\textbf{Cut Number}}}}%
      \put(3874,154){\makebox(0,0){\strut{}\textbf{T}}}%
      \put(178,4709){\makebox(0,0){\strut{}\textbf{c)}}}%
    }%
    \gplgaddtomacro\gplfronttext{%
      \put(5882,4426){\makebox(0,0)[r]{\strut{}$Hx$ [111] Ice Long Loop}}%
      \csname LTb\endcsname%
      \put(5882,4206){\makebox(0,0)[r]{\strut{}$Hy$ [111] Ice Long Loop}}%
    }%
    \gplgaddtomacro\gplbacktext{%
      \csname LTb\endcsname%
      \put(898,3355){\makebox(0,0)[r]{\strut{}}}%
      \put(898,3530){\makebox(0,0)[r]{\strut{}}}%
      \put(898,3705){\makebox(0,0)[r]{\strut{}}}%
      \put(898,3880){\makebox(0,0)[r]{\strut{}}}%
      \put(898,4055){\makebox(0,0)[r]{\strut{}}}%
      \put(898,4230){\makebox(0,0)[r]{\strut{}}}%
      \put(898,4405){\makebox(0,0)[r]{\strut{}}}%
      \put(1030,3135){\makebox(0,0){\strut{} 0}}%
      \put(2092,3135){\makebox(0,0){\strut{} 6}}%
      \put(3154,3135){\makebox(0,0){\strut{} 12}}%
    }%
    \gplgaddtomacro\gplfronttext{%
      \put(2167,4232){\makebox(0,0)[r]{\strut{}$Hx$ T=0.1}}%
      \csname LTb\endcsname%
      \put(2167,4012){\makebox(0,0)[r]{\strut{}$Hy$ T=0.1}}%
    }%
    \gplgaddtomacro\gplbacktext{%
      \csname LTb\endcsname%
      \put(4506,1070){\makebox(0,0)[r]{\strut{}}}%
      \put(4506,1201){\makebox(0,0)[r]{\strut{}}}%
      \put(4506,1332){\makebox(0,0)[r]{\strut{}}}%
      \put(4506,1464){\makebox(0,0)[r]{\strut{}}}%
      \put(4506,1595){\makebox(0,0)[r]{\strut{}}}%
      \put(4506,1726){\makebox(0,0)[r]{\strut{}}}%
      \put(4506,1857){\makebox(0,0)[r]{\strut{}}}%
      \put(4506,1988){\makebox(0,0)[r]{\strut{}}}%
      \put(4638,850){\makebox(0,0){\strut{} 0}}%
      \put(5664,850){\makebox(0,0){\strut{} 6}}%
      \put(6689,850){\makebox(0,0){\strut{} 12}}%
    }%
    \gplgaddtomacro\gplfronttext{%
      \put(5702,1881){\makebox(0,0)[r]{\strut{}$Hx$ T=10}}%
      \csname LTb\endcsname%
      \put(5702,1661){\makebox(0,0)[r]{\strut{}$Hy$ T=10}}%
    }%
    \gplbacktext
    \put(0,0){\includegraphics{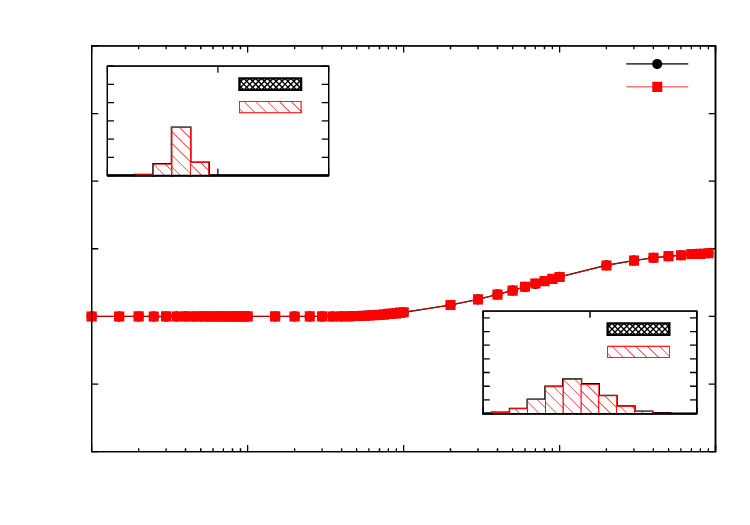}}%
    \gplfronttext
  \end{picture}%
\endgroup

%% file: Pseudo_Sus.tex
% GNUPLOT: LaTeX picture with Postscript
\begingroup
  \makeatletter
  \providecommand\color[2][]{%
    \GenericError{(gnuplot) \space\space\space\@spaces}{%
      Package color not loaded in conjunction with
      terminal option `colourtext'%
    }{See the gnuplot documentation for explanation.%
    }{Either use 'blacktext' in gnuplot or load the package
      color.sty in LaTeX.}%
    \renewcommand\color[2][]{}%
  }%
  \providecommand\includegraphics[2][]{%
    \GenericError{(gnuplot) \space\space\space\@spaces}{%
      Package graphicx or graphics not loaded%
    }{See the gnuplot documentation for explanation.%
    }{The gnuplot epslatex terminal needs graphicx.sty or graphics.sty.}%
    \renewcommand\includegraphics[2][]{}%
  }%
  \providecommand\rotatebox[2]{#2}%
  \@ifundefined{ifGPcolor}{%
    \newif\ifGPcolor
    \GPcolortrue
  }{}%
  \@ifundefined{ifGPblacktext}{%
    \newif\ifGPblacktext
    \GPblacktexttrue
  }{}%
  % define a \g@addto@macro without @ in the name:
  \let\gplgaddtomacro\g@addto@macro
  % define empty templates for all commands taking text:
  \gdef\gplbacktext{}%
  \gdef\gplfronttext{}%
  \makeatother
  \ifGPblacktext
    % no textcolor at all
    \def\colorrgb#1{}%
    \def\colorgray#1{}%
  \else
    % gray or color?
    \ifGPcolor
      \def\colorrgb#1{\color[rgb]{#1}}%
      \def\colorgray#1{\color[gray]{#1}}%
      \expandafter\def\csname LTw\endcsname{\color{white}}%
      \expandafter\def\csname LTb\endcsname{\color{black}}%
      \expandafter\def\csname LTa\endcsname{\color{black}}%
      \expandafter\def\csname LT0\endcsname{\color[rgb]{1,0,0}}%
      \expandafter\def\csname LT1\endcsname{\color[rgb]{0,1,0}}%
      \expandafter\def\csname LT2\endcsname{\color[rgb]{0,0,1}}%
      \expandafter\def\csname LT3\endcsname{\color[rgb]{1,0,1}}%
      \expandafter\def\csname LT4\endcsname{\color[rgb]{0,1,1}}%
      \expandafter\def\csname LT5\endcsname{\color[rgb]{1,1,0}}%
      \expandafter\def\csname LT6\endcsname{\color[rgb]{0,0,0}}%
      \expandafter\def\csname LT7\endcsname{\color[rgb]{1,0.3,0}}%
      \expandafter\def\csname LT8\endcsname{\color[rgb]{0.5,0.5,0.5}}%
    \else
      % gray
      \def\colorrgb#1{\color{black}}%
      \def\colorgray#1{\color[gray]{#1}}%
      \expandafter\def\csname LTw\endcsname{\color{white}}%
      \expandafter\def\csname LTb\endcsname{\color{black}}%
      \expandafter\def\csname LTa\endcsname{\color{black}}%
      \expandafter\def\csname LT0\endcsname{\color{black}}%
      \expandafter\def\csname LT1\endcsname{\color{black}}%
      \expandafter\def\csname LT2\endcsname{\color{black}}%
      \expandafter\def\csname LT3\endcsname{\color{black}}%
      \expandafter\def\csname LT4\endcsname{\color{black}}%
      \expandafter\def\csname LT5\endcsname{\color{black}}%
      \expandafter\def\csname LT6\endcsname{\color{black}}%
      \expandafter\def\csname LT7\endcsname{\color{black}}%
      \expandafter\def\csname LT8\endcsname{\color{black}}%
    \fi
  \fi
  \setlength{\unitlength}{0.0500bp}%
  \begin{picture}(6480.00,4536.00)%
    \gplgaddtomacro\gplbacktext{%
      \csname LTb\endcsname%
      \put(880,704){\makebox(0,0)[r]{\strut{} 0}}%
      \put(880,1382){\makebox(0,0)[r]{\strut{} 0.2}}%
      \put(880,2060){\makebox(0,0)[r]{\strut{} 0.4}}%
      \put(880,2739){\makebox(0,0)[r]{\strut{} 0.6}}%
      \put(880,3417){\makebox(0,0)[r]{\strut{} 0.8}}%
      \put(880,4095){\makebox(0,0)[r]{\strut{} 1}}%
      \put(1012,484){\makebox(0,0){\strut{} 0.01}}%
      \put(2296,484){\makebox(0,0){\strut{} 0.1}}%
      \put(3581,484){\makebox(0,0){\strut{} 1}}%
      \put(4865,484){\makebox(0,0){\strut{} 10}}%
      \put(6149,484){\makebox(0,0){\strut{} 100}}%
      \put(308,2399){\rotatebox{-270}{\makebox(0,0){\strut{}\textbf{T$\chi_p$}}}}%
      \put(3580,154){\makebox(0,0){\strut{}\textbf{T}}}%
      \put(280,4205){\makebox(0,0){\strut{}\textbf{a)}}}%
    }%
    \gplgaddtomacro\gplfronttext{%
      \csname LTb\endcsname%
      \put(3520,3922){\makebox(0,0)[r]{\strut{}$[111]$ Long Loop}}%
      \csname LTb\endcsname%
      \put(3520,3702){\makebox(0,0)[r]{\strut{}$[111]$ Short Loop}}%
      \csname LTb\endcsname%
      \put(3520,3482){\makebox(0,0)[r]{\strut{}Will's Ice}}%
    }%
    \gplbacktext
    \put(0,0){\includegraphics{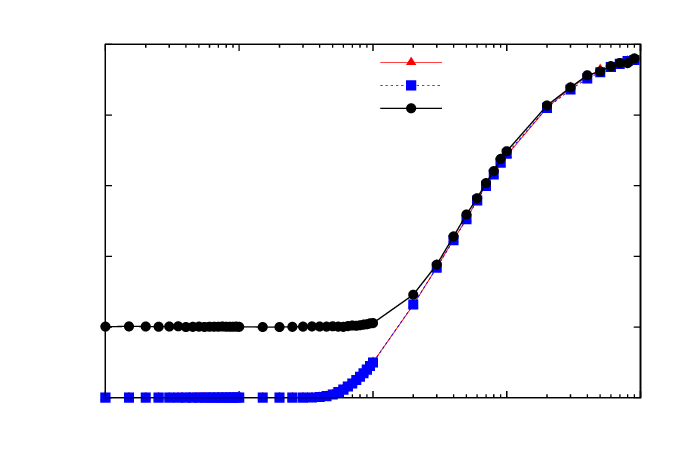}}%
    \gplfronttext
  \end{picture}%
\endgroup

%% file: Real_Sus.tex
% GNUPLOT: LaTeX picture with Postscript
\begingroup
  \makeatletter
  \providecommand\color[2][]{%
    \GenericError{(gnuplot) \space\space\space\@spaces}{%
      Package color not loaded in conjunction with
      terminal option `colourtext'%
    }{See the gnuplot documentation for explanation.%
    }{Either use 'blacktext' in gnuplot or load the package
      color.sty in LaTeX.}%
    \renewcommand\color[2][]{}%
  }%
  \providecommand\includegraphics[2][]{%
    \GenericError{(gnuplot) \space\space\space\@spaces}{%
      Package graphicx or graphics not loaded%
    }{See the gnuplot documentation for explanation.%
    }{The gnuplot epslatex terminal needs graphicx.sty or graphics.sty.}%
    \renewcommand\includegraphics[2][]{}%
  }%
  \providecommand\rotatebox[2]{#2}%
  \@ifundefined{ifGPcolor}{%
    \newif\ifGPcolor
    \GPcolortrue
  }{}%
  \@ifundefined{ifGPblacktext}{%
    \newif\ifGPblacktext
    \GPblacktexttrue
  }{}%
  % define a \g@addto@macro without @ in the name:
  \let\gplgaddtomacro\g@addto@macro
  % define empty templates for all commands taking text:
  \gdef\gplbacktext{}%
  \gdef\gplfronttext{}%
  \makeatother
  \ifGPblacktext
    % no textcolor at all
    \def\colorrgb#1{}%
    \def\colorgray#1{}%
  \else
    % gray or color?
    \ifGPcolor
      \def\colorrgb#1{\color[rgb]{#1}}%
      \def\colorgray#1{\color[gray]{#1}}%
      \expandafter\def\csname LTw\endcsname{\color{white}}%
      \expandafter\def\csname LTb\endcsname{\color{black}}%
      \expandafter\def\csname LTa\endcsname{\color{black}}%
      \expandafter\def\csname LT0\endcsname{\color[rgb]{1,0,0}}%
      \expandafter\def\csname LT1\endcsname{\color[rgb]{0,1,0}}%
      \expandafter\def\csname LT2\endcsname{\color[rgb]{0,0,1}}%
      \expandafter\def\csname LT3\endcsname{\color[rgb]{1,0,1}}%
      \expandafter\def\csname LT4\endcsname{\color[rgb]{0,1,1}}%
      \expandafter\def\csname LT5\endcsname{\color[rgb]{1,1,0}}%
      \expandafter\def\csname LT6\endcsname{\color[rgb]{0,0,0}}%
      \expandafter\def\csname LT7\endcsname{\color[rgb]{1,0.3,0}}%
      \expandafter\def\csname LT8\endcsname{\color[rgb]{0.5,0.5,0.5}}%
    \else
      % gray
      \def\colorrgb#1{\color{black}}%
      \def\colorgray#1{\color[gray]{#1}}%
      \expandafter\def\csname LTw\endcsname{\color{white}}%
      \expandafter\def\csname LTb\endcsname{\color{black}}%
      \expandafter\def\csname LTa\endcsname{\color{black}}%
      \expandafter\def\csname LT0\endcsname{\color{black}}%
      \expandafter\def\csname LT1\endcsname{\color{black}}%
      \expandafter\def\csname LT2\endcsname{\color{black}}%
      \expandafter\def\csname LT3\endcsname{\color{black}}%
      \expandafter\def\csname LT4\endcsname{\color{black}}%
      \expandafter\def\csname LT5\endcsname{\color{black}}%
      \expandafter\def\csname LT6\endcsname{\color{black}}%
      \expandafter\def\csname LT7\endcsname{\color{black}}%
      \expandafter\def\csname LT8\endcsname{\color{black}}%
    \fi
  \fi
  \setlength{\unitlength}{0.0500bp}%
  \begin{picture}(6480.00,4536.00)%
    \gplgaddtomacro\gplbacktext{%
      \csname LTb\endcsname%
      \put(880,704){\makebox(0,0)[r]{\strut{} 0}}%
      \put(880,1318){\makebox(0,0)[r]{\strut{} 0.5}}%
      \put(880,1933){\makebox(0,0)[r]{\strut{} 1}}%
      \put(880,2547){\makebox(0,0)[r]{\strut{} 1.5}}%
      \put(880,3161){\makebox(0,0)[r]{\strut{} 2}}%
      \put(880,3776){\makebox(0,0)[r]{\strut{} 2.5}}%
      \put(1012,484){\makebox(0,0){\strut{} 0.01}}%
      \put(2296,484){\makebox(0,0){\strut{} 0.1}}%
      \put(3581,484){\makebox(0,0){\strut{} 1}}%
      \put(4865,484){\makebox(0,0){\strut{} 10}}%
      \put(6149,484){\makebox(0,0){\strut{} 100}}%
      \put(308,2399){\rotatebox{-270}{\makebox(0,0){\strut{}\textbf{T$\chi_r$}}}}%
      \put(3580,154){\makebox(0,0){\strut{}\textbf{T}}}%
      \put(280,4205){\makebox(0,0){\strut{}\textbf{b)}}}%
    }%
    \gplgaddtomacro\gplfronttext{%
      \csname LTb\endcsname%
      \put(5162,3922){\makebox(0,0)[r]{\strut{}$[111]$ Long Loop}}%
      \csname LTb\endcsname%
      \put(5162,3702){\makebox(0,0)[r]{\strut{}$[111]$ Short Loop}}%
      \csname LTb\endcsname%
      \put(5162,3482){\makebox(0,0)[r]{\strut{}Will's Ice}}%
    }%
    \gplbacktext
    \put(0,0){\includegraphics{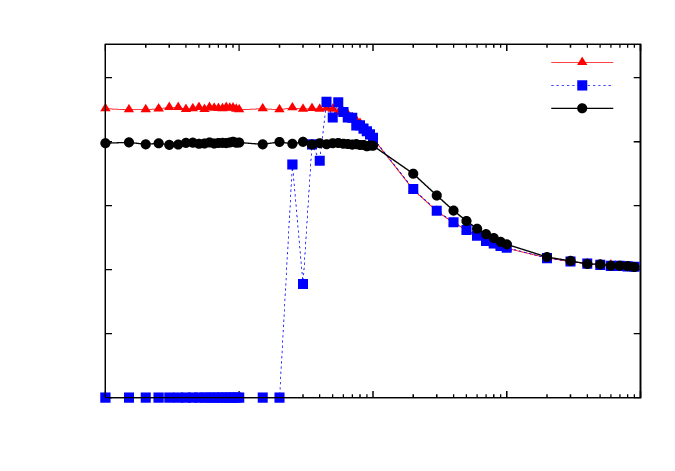}}%
    \gplfronttext
  \end{picture}%
\endgroup

%% file: RS_long.tex
% GNUPLOT: LaTeX picture with Postscript
\begingroup
  \makeatletter
  \providecommand\color[2][]{%
    \GenericError{(gnuplot) \space\space\space\@spaces}{%
      Package color not loaded in conjunction with
      terminal option `colourtext'%
    }{See the gnuplot documentation for explanation.%
    }{Either use 'blacktext' in gnuplot or load the package
      color.sty in LaTeX.}%
    \renewcommand\color[2][]{}%
  }%
  \providecommand\includegraphics[2][]{%
    \GenericError{(gnuplot) \space\space\space\@spaces}{%
      Package graphicx or graphics not loaded%
    }{See the gnuplot documentation for explanation.%
    }{The gnuplot epslatex terminal needs graphicx.sty or graphics.sty.}%
    \renewcommand\includegraphics[2][]{}%
  }%
  \providecommand\rotatebox[2]{#2}%
  \@ifundefined{ifGPcolor}{%
    \newif\ifGPcolor
    \GPcolortrue
  }{}%
  \@ifundefined{ifGPblacktext}{%
    \newif\ifGPblacktext
    \GPblacktexttrue
  }{}%
  % define a \g@addto@macro without @ in the name:
  \let\gplgaddtomacro\g@addto@macro
  % define empty templates for all commands taking text:
  \gdef\gplbacktext{}%
  \gdef\gplfronttext{}%
  \makeatother
  \ifGPblacktext
    % no textcolor at all
    \def\colorrgb#1{}%
    \def\colorgray#1{}%
  \else
    % gray or color?
    \ifGPcolor
      \def\colorrgb#1{\color[rgb]{#1}}%
      \def\colorgray#1{\color[gray]{#1}}%
      \expandafter\def\csname LTw\endcsname{\color{white}}%
      \expandafter\def\csname LTb\endcsname{\color{black}}%
      \expandafter\def\csname LTa\endcsname{\color{black}}%
      \expandafter\def\csname LT0\endcsname{\color[rgb]{1,0,0}}%
      \expandafter\def\csname LT1\endcsname{\color[rgb]{0,1,0}}%
      \expandafter\def\csname LT2\endcsname{\color[rgb]{0,0,1}}%
      \expandafter\def\csname LT3\endcsname{\color[rgb]{1,0,1}}%
      \expandafter\def\csname LT4\endcsname{\color[rgb]{0,1,1}}%
      \expandafter\def\csname LT5\endcsname{\color[rgb]{1,1,0}}%
      \expandafter\def\csname LT6\endcsname{\color[rgb]{0,0,0}}%
      \expandafter\def\csname LT7\endcsname{\color[rgb]{1,0.3,0}}%
      \expandafter\def\csname LT8\endcsname{\color[rgb]{0.5,0.5,0.5}}%
    \else
      % gray
      \def\colorrgb#1{\color{black}}%
      \def\colorgray#1{\color[gray]{#1}}%
      \expandafter\def\csname LTw\endcsname{\color{white}}%
      \expandafter\def\csname LTb\endcsname{\color{black}}%
      \expandafter\def\csname LTa\endcsname{\color{black}}%
      \expandafter\def\csname LT0\endcsname{\color{black}}%
      \expandafter\def\csname LT1\endcsname{\color{black}}%
      \expandafter\def\csname LT2\endcsname{\color{black}}%
      \expandafter\def\csname LT3\endcsname{\color{black}}%
      \expandafter\def\csname LT4\endcsname{\color{black}}%
      \expandafter\def\csname LT5\endcsname{\color{black}}%
      \expandafter\def\csname LT6\endcsname{\color{black}}%
      \expandafter\def\csname LT7\endcsname{\color{black}}%
      \expandafter\def\csname LT8\endcsname{\color{black}}%
    \fi
  \fi
  \setlength{\unitlength}{0.0500bp}%
  \begin{picture}(5760.00,4032.00)%
    \gplgaddtomacro\gplbacktext{%
      \csname LTb\endcsname%
      \put(1078,704){\makebox(0,0)[r]{\strut{} 1.4}}%
      \put(1078,1216){\makebox(0,0)[r]{\strut{} 1.6}}%
      \put(1078,1728){\makebox(0,0)[r]{\strut{} 1.8}}%
      \put(1078,2240){\makebox(0,0)[r]{\strut{} 2}}%
      \put(1078,2751){\makebox(0,0)[r]{\strut{} 2.2}}%
      \put(1078,3263){\makebox(0,0)[r]{\strut{} 2.4}}%
      \put(1078,3775){\makebox(0,0)[r]{\strut{} 2.6}}%
      \put(1210,484){\makebox(0,0){\strut{} 0}}%
      \put(2054,484){\makebox(0,0){\strut{} 0.1}}%
      \put(2898,484){\makebox(0,0){\strut{} 0.2}}%
      \put(3741,484){\makebox(0,0){\strut{} 0.3}}%
      \put(4585,484){\makebox(0,0){\strut{} 0.4}}%
      \put(5429,484){\makebox(0,0){\strut{} 0.5}}%
      \put(308,2367){\rotatebox{-270}{\makebox(0,0){\strut{}\textbf{T$\chi_r$}}}}%
      \put(3319,154){\makebox(0,0){\strut{}\textbf{1/L}}}%
    }%
    \gplgaddtomacro\gplfronttext{%
      \csname LTb\endcsname%
      \put(4442,3858){\makebox(0,0)[r]{\strut{}$\chi_r$ at  T=0.01}}%
    }%
    \gplbacktext
    \put(0,0){\includegraphics{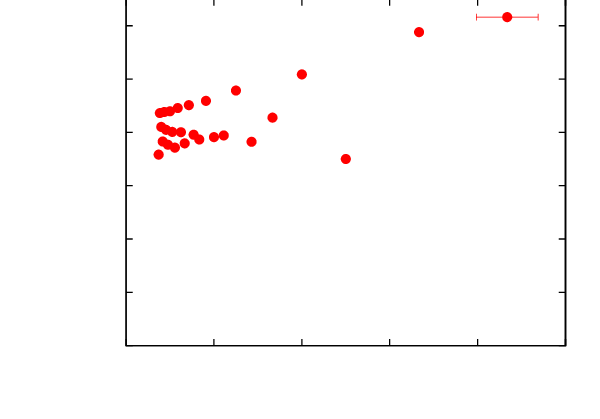}}%
    \gplfronttext
  \end{picture}%
\endgroup

%% file: RS_subset.tex
% GNUPLOT: LaTeX picture with Postscript
\begingroup
  \makeatletter
  \providecommand\color[2][]{%
    \GenericError{(gnuplot) \space\space\space\@spaces}{%
      Package color not loaded in conjunction with
      terminal option `colourtext'%
    }{See the gnuplot documentation for explanation.%
    }{Either use 'blacktext' in gnuplot or load the package
      color.sty in LaTeX.}%
    \renewcommand\color[2][]{}%
  }%
  \providecommand\includegraphics[2][]{%
    \GenericError{(gnuplot) \space\space\space\@spaces}{%
      Package graphicx or graphics not loaded%
    }{See the gnuplot documentation for explanation.%
    }{The gnuplot epslatex terminal needs graphicx.sty or graphics.sty.}%
    \renewcommand\includegraphics[2][]{}%
  }%
  \providecommand\rotatebox[2]{#2}%
  \@ifundefined{ifGPcolor}{%
    \newif\ifGPcolor
    \GPcolortrue
  }{}%
  \@ifundefined{ifGPblacktext}{%
    \newif\ifGPblacktext
    \GPblacktexttrue
  }{}%
  % define a \g@addto@macro without @ in the name:
  \let\gplgaddtomacro\g@addto@macro
  % define empty templates for all commands taking text:
  \gdef\gplbacktext{}%
  \gdef\gplfronttext{}%
  \makeatother
  \ifGPblacktext
    % no textcolor at all
    \def\colorrgb#1{}%
    \def\colorgray#1{}%
  \else
    % gray or color?
    \ifGPcolor
      \def\colorrgb#1{\color[rgb]{#1}}%
      \def\colorgray#1{\color[gray]{#1}}%
      \expandafter\def\csname LTw\endcsname{\color{white}}%
      \expandafter\def\csname LTb\endcsname{\color{black}}%
      \expandafter\def\csname LTa\endcsname{\color{black}}%
      \expandafter\def\csname LT0\endcsname{\color[rgb]{1,0,0}}%
      \expandafter\def\csname LT1\endcsname{\color[rgb]{0,1,0}}%
      \expandafter\def\csname LT2\endcsname{\color[rgb]{0,0,1}}%
      \expandafter\def\csname LT3\endcsname{\color[rgb]{1,0,1}}%
      \expandafter\def\csname LT4\endcsname{\color[rgb]{0,1,1}}%
      \expandafter\def\csname LT5\endcsname{\color[rgb]{1,1,0}}%
      \expandafter\def\csname LT6\endcsname{\color[rgb]{0,0,0}}%
      \expandafter\def\csname LT7\endcsname{\color[rgb]{1,0.3,0}}%
      \expandafter\def\csname LT8\endcsname{\color[rgb]{0.5,0.5,0.5}}%
    \else
      % gray
      \def\colorrgb#1{\color{black}}%
      \def\colorgray#1{\color[gray]{#1}}%
      \expandafter\def\csname LTw\endcsname{\color{white}}%
      \expandafter\def\csname LTb\endcsname{\color{black}}%
      \expandafter\def\csname LTa\endcsname{\color{black}}%
      \expandafter\def\csname LT0\endcsname{\color{black}}%
      \expandafter\def\csname LT1\endcsname{\color{black}}%
      \expandafter\def\csname LT2\endcsname{\color{black}}%
      \expandafter\def\csname LT3\endcsname{\color{black}}%
      \expandafter\def\csname LT4\endcsname{\color{black}}%
      \expandafter\def\csname LT5\endcsname{\color{black}}%
      \expandafter\def\csname LT6\endcsname{\color{black}}%
      \expandafter\def\csname LT7\endcsname{\color{black}}%
      \expandafter\def\csname LT8\endcsname{\color{black}}%
    \fi
  \fi
  \setlength{\unitlength}{0.0500bp}%
  \begin{picture}(6480.00,4536.00)%
    \gplgaddtomacro\gplbacktext{%
      \csname LTb\endcsname%
      \put(1078,704){\makebox(0,0)[r]{\strut{} 0}}%
      \put(1078,1343){\makebox(0,0)[r]{\strut{} 0.5}}%
      \put(1078,1981){\makebox(0,0)[r]{\strut{} 1}}%
      \put(1078,2620){\makebox(0,0)[r]{\strut{} 1.5}}%
      \put(1078,3258){\makebox(0,0)[r]{\strut{} 2}}%
      \put(1078,3897){\makebox(0,0)[r]{\strut{} 2.5}}%
      \put(1078,4535){\makebox(0,0)[r]{\strut{} 3}}%
      \put(1210,484){\makebox(0,0){\strut{} 0}}%
      \put(1958,484){\makebox(0,0){\strut{} 5}}%
      \put(2707,484){\makebox(0,0){\strut{} 10}}%
      \put(3455,484){\makebox(0,0){\strut{} 15}}%
      \put(4203,484){\makebox(0,0){\strut{} 20}}%
      \put(4952,484){\makebox(0,0){\strut{} 25}}%
      \put(5700,484){\makebox(0,0){\strut{} 30}}%
      \put(308,2619){\rotatebox{-270}{\makebox(0,0){\strut{}\textbf{T$\chi_r$}}}}%
      \put(3679,154){\makebox(0,0){\strut{}\textbf{$\ell$}}}%
    }%
    \gplgaddtomacro\gplfronttext{%
      \csname LTb\endcsname%
      \put(3190,4362){\makebox(0,0)[r]{\strut{}Loop Size = 6}}%
      \csname LTb\endcsname%
      \put(3190,4142){\makebox(0,0)[r]{\strut{}Loop Size = 18}}%
      \csname LTb\endcsname%
      \put(3190,3922){\makebox(0,0)[r]{\strut{}Loop Size = 32}}%
      \csname LTb\endcsname%
      \put(3190,3702){\makebox(0,0)[r]{\strut{}Loop Size = 63}}%
      \csname LTb\endcsname%
      \put(3190,3482){\makebox(0,0)[r]{\strut{}All Loop Sizes}}%
    }%
    \gplbacktext
    \put(0,0){\includegraphics{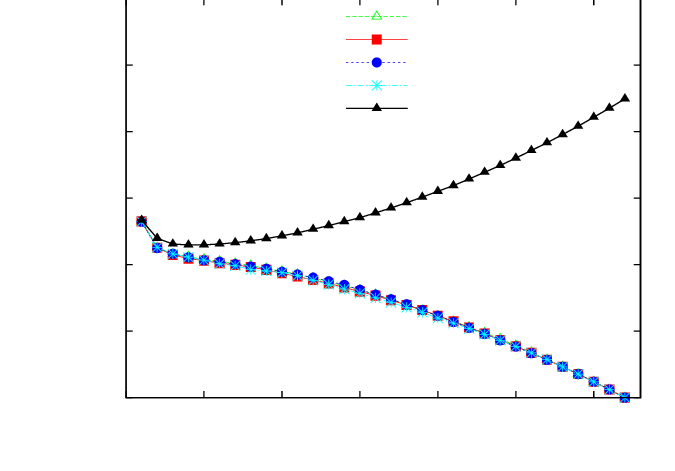}}%
    \gplfronttext
  \end{picture}%
\endgroup